\documentclass[10pt,aps,prl,twocolumn,notitlepage,nofootinbib,longbibliography,floatfix,superscriptaddress]{revtex4-1}

\usepackage[T1]{fontenc}
\usepackage{graphicx} 
\usepackage{bm}        
\usepackage{amssymb}  
\usepackage{mathtools}
\usepackage{amsmath}
\usepackage{amsthm}
\usepackage[colorlinks=true,linkcolor=blue,citecolor=blue,urlcolor=blue,plainpages=false,pdfpagelabels]{hyperref}
\usepackage{color,xcolor,colortbl}
\usepackage{multirow}
\usepackage{bbm}
\usepackage{float}
\usepackage{array}
\usepackage{enumitem}
\usepackage{transparent}
\usepackage{verbatim}
\usepackage[normalem]{ulem}
\usepackage{cancel}

\newcommand{\bra}[1]{\langle #1|}
\newcommand{\ket}[1]{|#1\rangle}
\newcommand{\dyad}[1]{\ket{#1}\!\bra{#1}}

\newcommand{\Tr}{\text{Tr}}

\newcommand{\norm}[1]{\lVert#1\rVert}

\newcommand{\e}{\text{e}}
\DeclareMathOperator*{\argmax}{arg\,max}

\makeatletter
\g@addto@macro\bfseries{\boldmath}
\makeatother

\theoremstyle{plain}%

\newtheorem*{theorem*}{Theorem}

\allowdisplaybreaks

\AtBeginDocument{%
    \newwrite\bibnotes
    \def\bibnotesext{Notes.bib}
    \immediate\openout\bibnotes=\jobname\bibnotesext
    \immediate\write\bibnotes{@CONTROL{REVTEX41Control}}
    \immediate\write\bibnotes{@CONTROL{%
    apsrev41Control,author="08",editor="1",pages="1",title="0",year="1"}}
     \if@filesw
     \immediate\write\@auxout{\string\citation{apsrev41Control}}%
    \fi
}%

\begin{document}

\widetext

\title{Spooky action at a global distance: analysis of space-based entanglement distribution for the quantum internet}

\author{Sumeet Khatri}\thanks{Equal contribution} \email{skhatr5@lsu.edu}\affiliation{Hearne Institute for Theoretical Physics, Department of Physics and Astronomy, Louisiana State University, Baton Rouge, Louisiana, 70803, USA}

\author{Anthony J. Brady}\thanks{Equal contribution}\email{abrady6@lsu.edu}\affiliation{Hearne Institute for Theoretical Physics, Department of Physics and Astronomy, Louisiana State University, Baton Rouge, Louisiana, 70803, USA}

\author{Ren\'{e}e A. Desporte}\affiliation{Hearne Institute for Theoretical Physics, Department of Physics and Astronomy, Louisiana State University, Baton Rouge, Louisiana, 70803, USA}

\author{Manon P. Bart}\affiliation{Hearne Institute for Theoretical Physics, Department of Physics and Astronomy, Louisiana State University, Baton Rouge, Louisiana, 70803, USA}

\author{Jonathan P. Dowling} \affiliation{Hearne Institute for Theoretical Physics, Department of Physics and Astronomy, Louisiana State University, Baton Rouge, Louisiana, 70803, USA}\affiliation{National Institute of Information and Communications Technology, 4-2-1, Nukui-Kitamachi, Koganei, Tokyo 184-8795, Japan}\affiliation{NYU-ECNU Institute of Physics at NYU Shanghai, Shanghai 200062, China}\affiliation{CAS-Alibaba Quantum Computing Laboratory, USTC, Shanghai 201315, China}

\date{\today}

\begin{abstract}
	
	Recent experimental breakthroughs in satellite quantum communications have opened up the possibility of creating a global quantum internet using satellite links. This approach appears to be particularly viable in the near term, due to the lower attenuation of optical signals from satellite to ground, and due to the currently short coherence times of quantum memories. The latter prevents ground-based entanglement distribution using atmospheric or optical-fiber links at high rates over long distances. In this work, we propose a global-scale quantum internet consisting of a constellation of orbiting satellites that provides a continuous, on-demand entanglement distribution service to ground stations. The satellites can also function as untrusted nodes for the purpose of long-distance quantum-key distribution. We develop a technique for determining optimal satellite configurations with continuous coverage that balances both the total number of satellites and entanglement-distribution rates. Using this technique, we determine various optimal satellite configurations for a polar-orbit constellation, and we analyze the resulting satellite-to-ground loss and achievable entanglement-distribution rates for multiple ground station configurations. We also provide a comparison between these entanglement-distribution rates and the rates of ground-based quantum repeater schemes. Overall, our work provides the theoretical tools and the experimental guidance needed to make a satellite-based global quantum internet a reality.

\end{abstract}

\maketitle


\section{Introduction}

	One of the most remarkable applications of quantum mechanics is the ability to perform secure communication via quantum-key distribution (QKD) \cite{BB84,Eke91,GRG+02,SBPC+09}. While current global communication systems rely on computational security and are breakable with a quantum computer \cite{Shor94,Shor97,MVZJ18}, QKD offers, in principle, unconditional (information-theoretic) security even against adversaries with a quantum computer. With several metropolitan-scale QKD systems already in place \cite{PPA+09,CWL+10,MP10,SLB+11,SFI+11,WCY+14,BLL+18,ZXCPP18}, and with the development of quantum computers proceeding at a steady pace \cite{LSB+19,BCMS19,arute2019quantum}, the time is right to begin transitioning to a global quantum communications network before full-scale quantum computers render current communication systems defenseless \cite{Mos15,GM18,MP19}. In addition to QKD, a global quantum communications network, or quantum internet \cite{Kim08,Sim17,Cast18,Wehn+18,Dowling_book2}, would allow for the execution of other quantum-information-processing tasks, such as quantum teleportation \cite{BBC+93,BFK00}, quantum clock synchronization \cite{JADW00,UD02,DB+18}, distributed quantum computation \cite{CEHM99}, and distributed quantum metrology and sensing \cite{DRC17,ZZS18,XZCZ19}.
	
	Building the quantum internet is a major experimental challenge. All of the aforementioned tasks make use of shared entanglement between distant locations on the earth, which is typically distributed using single-photonic qubits sent through either the atmosphere or optical fibers. These schemes require reliable single-photon sources, quantum memories with high coherence times, and quantum gate operations with low error. It is well known that optical signals transmitted through either the atmosphere or optical fibers undergo an exponential decrease in the transmission success probability with distance \cite{SveltoBook,KJK_book}. Quantum repeaters \cite{BDC98,DBC99,SSR+11} have been proposed to overcome this exponential loss by dividing the transmission line into smaller segments along which errors and loss can be corrected using entanglement swapping \cite{BBC+93,ZZH93} and entanglement purification \cite{BBP96,DAR96,BDD96}. Several theoretical proposals for quantum repeater schemes have been made (see Refs. \cite{SSR+11,Terhal15,MLK+16} and references therein); however, many of these proposals have resource requirements that are currently unattainable. Furthermore, experimental demonstrations performed so far have been limited \cite{Hump+18,Kalb+17,Ritter+12} and do not scale to the distances needed to realize a global-scale quantum internet.
	
	\begin{figure*}
		\centering 
		\includegraphics[scale=.8]{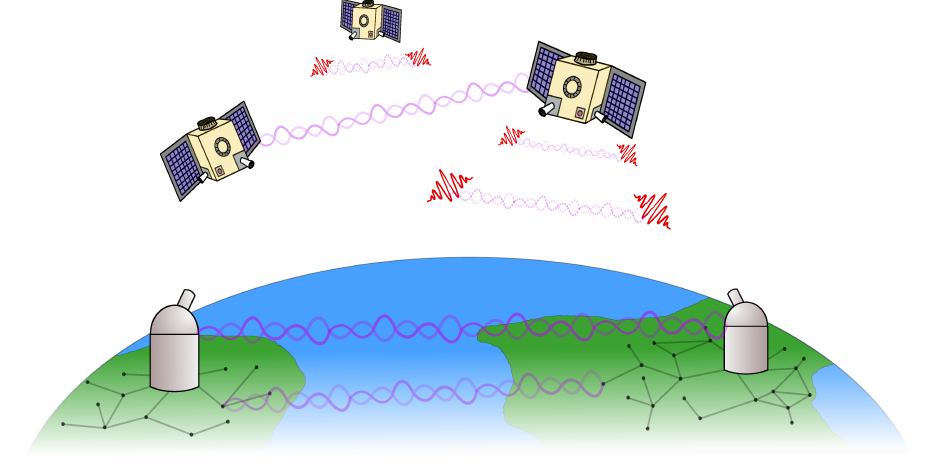}
		\caption{A hybrid global quantum communications network. A satellite constellation distributes entangled photon pairs (red wave packets; entanglement depicted by wavy lines) to distant ground stations (observatories) that host multimode quantum memories for storage \cite{Simon+07,Sinclair+14,Yang+18}. These stations act as hubs that connect to local nodes (black dots) via fiber-optic or atmospheric links. Using these nearest-neighbor entangled links, via entanglement swapping, two distant nodes can share entanglement. Note that this architecture can support inter-satellite entanglement links as well, which is useful for exploring fundamental physics \cite{Bruschi+14}, and for forming an international time standard \cite{KKBJ+14}.} \label{fig-hybrid_network}
	\end{figure*}
	
	Satellites have been recognized as one of the best methods for achieving global-scale quantum communication with current or near-term resources \cite{AJP+03,JH13,BAL17,Sim17,Cubesat2017,Nanobob2018}. Using satellites is advantageous due to the fact that the majority of the optical path traversed by an entangled photon pair is in free space, resulting in lower loss compared to ground-based entanglement distribution over atmospheric or fiber-optic links. Satellites can also be used to implement long-distance QKD with untrusted nodes, which is missing from most current implementations of long-distance QKD due to the lack of a quantum repeater. A satellite-based approach also allows for the possibility to use quantum strategies for tasks such as establishing a robust and secure international time scale via a quantum network of clocks \cite{KKBJ+14}, extending the baseline of telescopes for improved astronomical imaging \cite{GJC12, 1KBDL18, 2KBDL18}, and exploring fundamental physics \cite{RJ+12, Bruschi+14}. 
	
	Several proposals for satellite-based quantum networks have been made that use satellite-to-ground transmission, ground-to-satellite transmission, or both \cite{AJP+03,Bon09,ESH+12,BMH+13,BBM+15,TCT+16,Bedington2016nanosatellite,Cubesat2017,Nanobob2018,HMG18,HMG19,VLBKL19}. Recent experiments \cite{TCT+16,LYL+2017,YCL17,LCL+17,TCF+17,Ren17satteleport,LCH+18,CCD+18} (see also Ref.~\cite{LV19} for a review) between a handful of nodes opens up the possibility of building a global-scale quantum internet using satellites. As shown in Fig.~\ref{fig-hybrid_network}, this means having a constellation of orbiting satellites that transmit either bipartite or multipartite entanglement to ground stations. These ground stations can act as hubs that then distribute entanglement to neighboring ground stations via short ground-based links. In order to successfully implement such a global-scale satellite-based quantum internet, many factors must be taken into account, such as economics, current technology, resource availability, and performance requirements. Ideally, the satellite network should have continuous global coverage and provide entanglement on demand at a reasonably high rate between any two distant points on earth. Given this performance requirement, important questions related to economics and resources arise, such as: How many satellites are needed for continuous global coverage? At what altitude should the satellites be placed? What entanglement-distribution rates are possible between any points on earth, and how do these rates compare to those that can be achieved using ground-based quantum repeater setups?
	
	In this work, we address these questions by analyzing a global-scale quantum internet architecture in which satellites arranged in a constellation of polar orbits (see Fig.~\ref{fig-sat_architecture}) act as entanglement sources that distribute entangled photon pairs to ground stations. The nearest-neighbor entangled links can then be extended via entanglement swapping to obtain shared entanglement over longer distances. We start by determining the required number of satellites for such a network to have continuous global coverage. Since satellites are a costly resource, continuous global coverage should be achieved with as few satellites as possible. To that end, our first contribution is to define a figure of merit that allows us to investigate the trade-off between the number of satellites, their altitude, the average loss over a 24-hour period, and the average entanglement-distribution rates. By running simulations in order to optimize our figure of merit, we obtain one of our main results, which is the optimal number of satellites needed for continuous global coverage, as well as the optimal altitude at which the satellites should be placed such that the average loss is below a certain threshold. We then compare the resulting entanglement-distribution rates to those obtained via a ground-based entanglement distribution scheme assisted by quantum repeaters. This leads to another key result of our work, which is that the satellite-based scheme (without quantum repeaters) can outperform ground-based quantum repeater schemes in certain cases. We also consider entanglement distribution to major global cities over intercontinental distances. The key result here is that, with a constellation of 400 satellites, entanglement distribution at a reasonably high rate is not possible beyond approximately 7500~km.
	
	We remark that our approach is similar to the approach taken in Ref.~\cite{BBM+15}, in which ground stations are placed only on the equator and there is a single ring of satellites in an equatorial orbit around the earth. Our work goes beyond this by considering a genuine network scenario in which multiple ground stations are placed arbitrarily on the earth and there is a constellation of satellites in polar rather than equatorial orbits, as shown in Fig.~\ref{fig-sat_architecture}. Furthermore, while prior work has considered satellite constellations for entanglement distribution \cite{VLBKL19,MLL+20}, to our knowledge, the type of dynamic quantum network simulation with satellite constellations that we consider, along with optimization over different constellation configurations, has not been previously studied.
	
	We expect the results of this work to serve as a guide for building a global-scale quantum internet, both in terms of the number of satellites needed as well as the expected performance of the network. In particular, our results comparing satellite-based entanglement distribution to ground-based repeater-assisted entanglement distribution suggest that, at least in the near term, satellites are indeed the most viable approach to obtaining a global-scale quantum internet.

\section{Results}

\subsection{Network architecture}\label{sec-net_arch}

\begin{figure}
		\centering
		\includegraphics[scale=0.4]{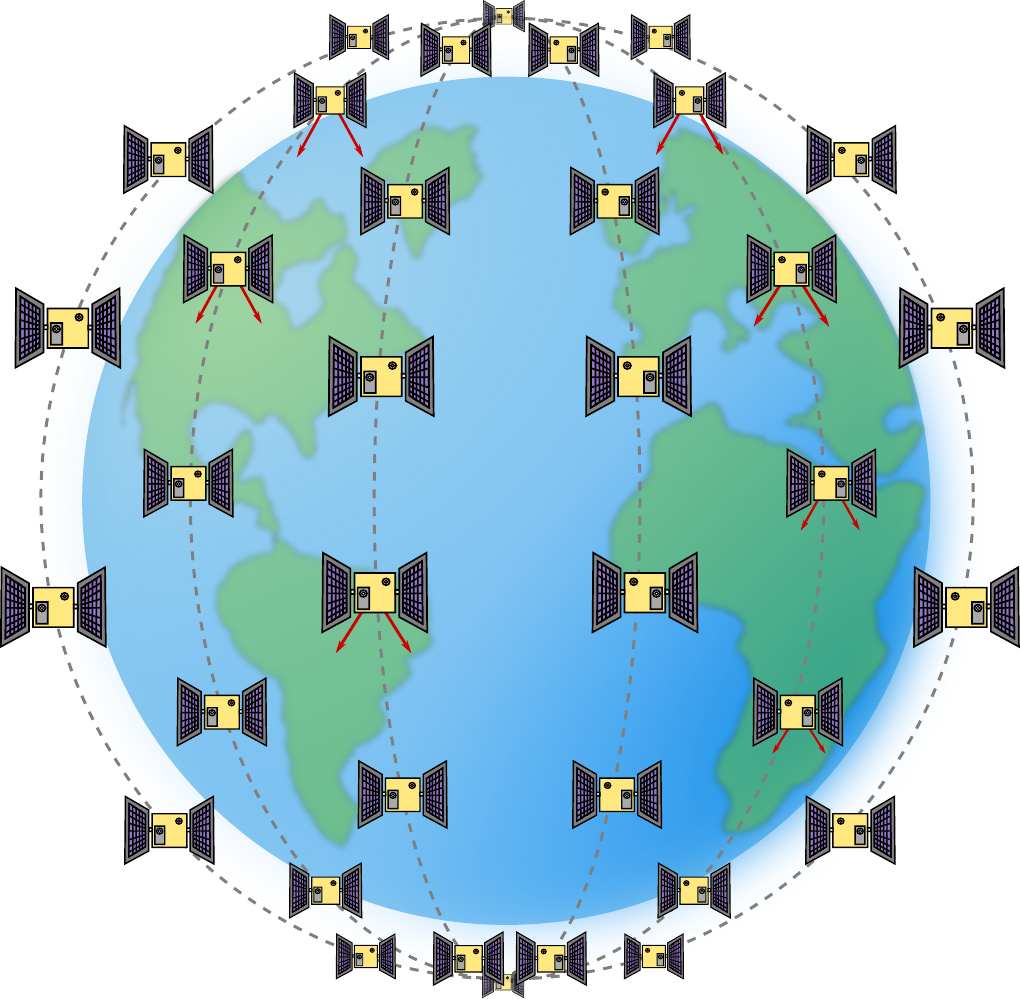}
		\caption{Our proposed satellite-based quantum network. We allow for $N_R$ equally-spaced rings of satellites. Within each ring, we allow for $N_S$ satellites in polar orbits.}\label{fig-sat_architecture}
	\end{figure}

	Our proposed satellite network architecture is illustrated in Fig. \ref{fig-sat_architecture}. We consider $N_R$ equally spaced rings of satellites in polar orbits. We allow for $N_S$ equally-spaced satellites in each ring, so that there are $N_RN_S$ satellites in total, all of which are at the same altitude $h$. This type of satellite constellation falls into the general class of Walker star constellations \cite{Walker70}, and we consider it mainly for its simplicity, but also because this constellation is similar to the Iridium communications-satellite constellation \cite{Leopold92,PRFT99}. Prior works have examined various other types of satellite constellations for the purpose of continuous global coverage \cite{Luders61,Walker70,LA98,LFP98}. The recent Starlink constellation \cite{Handley18} is also being used to provide a global satellite-based internet service. Investigations of these other satellite constellation types, and comparisons between them in the context of a global quantum internet, is an interesting direction for future work.
	
	The satellites act as source stations that transmit pairs of entangled photons to line-of-sight ground stations for the purpose of establishing elementary entanglement links. The ground stations can act as quantum repeaters in this scheme---performing entanglement purification and entanglement swapping once the elementary links have been established. In this way, we execute long-distance entanglement distribution between ground stations. Note that we could alternatively use the satellites as quantum repeaters \cite{LKB20,GSH+20}, which would require uplinks. It has been shown in, e.g., Ref. \cite{BMH+13}, that uplinks are more lossy and lead to lower key rates for QKD. For this reason, we consider downlinks only. The photon sources on the satellites produce polarization-entangled photon pairs. State-of-the-art sources of entangled photons are capable of producing polarization-entangled photons on a chip with a fidelity up to 0.97~\cite{HKK+13,matsuda2012monolithically,Kang:16,KRR+17}. 

\subsection{Overview of simulations}

	We obtain our results by running several entanglement distribution simulations using the satellite network architecture illustrated in Fig.~\ref{fig-sat_architecture}. We consider as our baseline requirement that a satellite network should provide continuous coverage to two ground stations located on the equator. We thus start by running a 24-hour simulation with two ground stations at the equator separated by distances $d$ between 100~km and 5000~km, and satellite configurations ranging from 20 to 400 satellites at altitudes between 500~km and 10000~km. We choose ground distances starting from 100~km because 100~km is roughly the longest distance at which ground-based entanglement distribution can be successfully performed at a reasonable rate without quantum repeaters; see, e.g., Refs.~\cite{DTY+09,YRL+12,IMT+13,WJS+19}. Our choice of satellite altitudes encompasses both low earth orbits and medium earth orbits, which are the orbits currently being used for most satellite communications systems \cite{PRFT99,Handley18}.
	
	A satellite configuration is given by the number $N_R$ of satellite rings, the number $N_S$ of satellites per ring, and the altitude $h$ of the satellites.  Our requirement of continuous coverage means that both ground stations must be simultaneously in view of a satellite at all times. We also impose an additional requirement that, even when in view of both ground stations, the total transmission loss between a satellite and the ground station pair should not exceed 90 dB, in order to keep ebit rates above $1$ Hz. (See the Methods section for further simulation details.) Note that, based on the satellite constellations that we consider here, two ground stations at the equator is the worst-case scenario, in the sense that two ground stations at higher or lower latitudes will always have less satellite-to-ground loss on average (we show this in Fig.~\ref{fig-rates_two_ground_stations} below).

	For all of our simulations, we take into account attenuation due to the atmosphere; see the Methods section for a description of our loss model. However, we assume clear skies, hence no rain, haze, or cloud coverage in any area. Including these extra elements would introduce extra attenuation factors (see, e.g., Ref.~\cite[Section~2.1.1.4]{KJK_book} and Refs.~\cite{Vogel2017weather,LKB+19}), which would increase the overall satellite-to-ground transmission loss. See Refs.~\cite{MLL+20,PMD+20} for an analysis of satellite-to-ground quantum key distribution in a localized area that incorporates local weather conditions. We also point out that, especially in the daytime, background photons (e.g., from the sun) can reduce the fidelity of the distributed entangled pairs, because the receiver will collect those background photons in addition to the signal photons from the entanglement source. This source of background photons is perhaps the most difficult obstacle to continuous global coverage. Timing information, as well as information about the spectral and spatial profile of the signal, can help reduce the noise via filtering, but only to a certain extent (see, e.g., Refs.~\cite{LYL+2017, Ko+2018daylight}). Furthermore, because the probability to transmit single photons from satellite to ground is quite low, the communicating parties must ensure that the probability to collect background photons is even lower in order to ensure a high signal-to-noise ratio (SNR), and thus a high fidelity for the received quantum state. In the Methods section we show how the fidelity of the transmitted states is affected by spurious background photons.

\subsection{Optimal network configurations for global coverage}\label{sec-opt_configs}

	\begin{figure*}
		\centering
		\includegraphics[width=\textwidth]{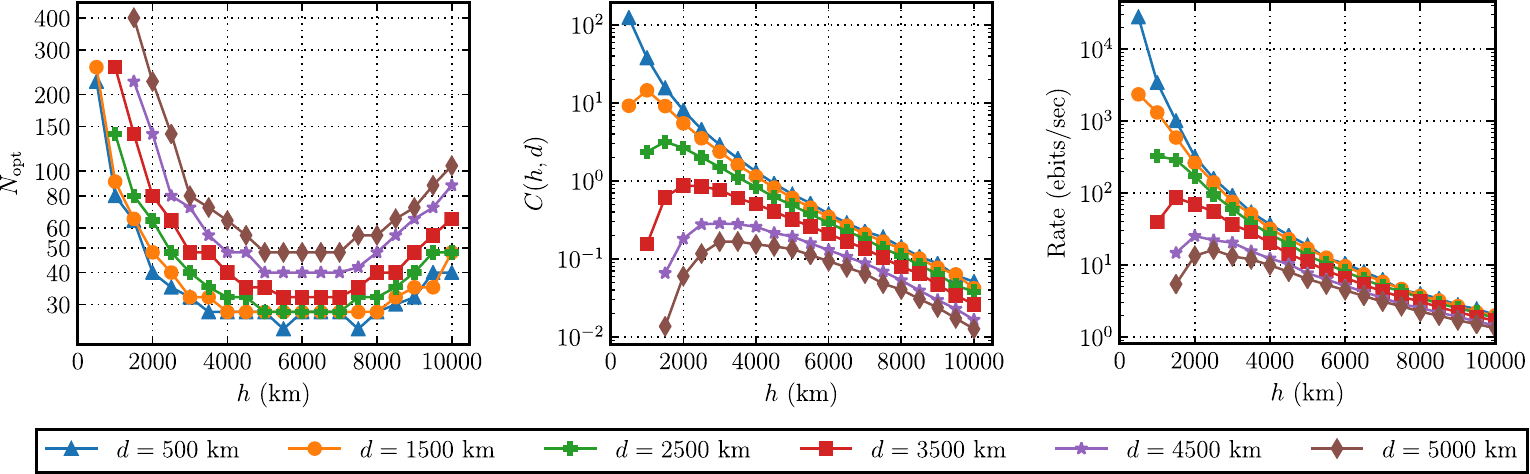}
		\caption{Simulation results for two ground stations at the equator separated by a distance $d$. (Left) Optimal number $N_{\text{opt}}(h,d)$ of satellites for continuous 24-hour coverage. (Center) Figure of merit in Eq.~\eqref{eq-cost_func_normalized} in units of ebits per second per satellite. Satellite configurations corresponding to the maxima of the curves are shown in Table~\ref{table_opt-params}. (Right) Entanglement-distribution rates corresponding to the points in the plot in the central panel. We assume a source rate of $R_{\text{source}}=10^9$ ebits per second~\cite{CLZ+18}.} \label{costfunc-graph1}
	\end{figure*}
	
	Given two ground stations separated by a distance $d$ and situated at the equator, along with a particular satellite constellation defined by $(N_R,N_S,h)$, as described above, how do we evaluate the performance of the given satellite constellation? Since satellites are currently an expensive resource, we would like to have as few satellites as possible in the network while still maintaining complete and continuous coverage. We could therefore take as our figure of merit the total number of satellites in the network. Specifically, given an altitude $h$ of the satellites and distance $d$ between the two ground stations, we define $N_{\text{opt}}(h,d)$ to be the minimum total number of satellites needed to have continuous 24-hour coverage for the two ground stations (see the Methods section for details). We could then minimize $N_{\text{opt}}(h,d)$ with respect to altitudes. On the other hand, we also want high entanglement distribution rates. We let $\overline{R}(N_R,N_S,h,d)$ denote the average entanglement-distribution rate over 24 hours for the satellite configuration given by $(N_R,N_S,h)$ and two ground stations at the equator separated by a distance $d$ (see the Methods section for the formal definition). The rate is calculated in a simple scenario without multimode transmission from the satellites and without multimode quantum memories at the ground stations. We could then take the quantity
	\begin{equation}\label{eq-opt_avg_ent_rate}
		\overline{R}^{\text{opt}}(h,d)\coloneqq\max_{N_R,N_S}\overline{R}(N_R,N_S,h,d)
	\end{equation}
	as our figure or merit, which is the average rate (in ebits per second) over a 24-hour period for a given altitude $h$ and a given distance $d$, where the optimization is over satellite configurations with a fixed $h$ such that there is continuous coverage for 24 hours and the loss at any time is less than 90~dB (see the Methods section for details). Now, as one might expect, with fewer satellites the average loss would increase, thus decreasing entanglement-distribution rates, while increasing the number of satellites would decrease the loss, hence increasing the average entanglement-distribution rate. In order to balance our two competing goals---minimizing the total number of satellites and also maximizing the average rate---we take as our figure of merit the ratio of the average entanglement-distribution rate to the total number of satellites:
	\begin{equation}\label{eq-cost_function}
		c(N_R,N_S,h,d)\coloneqq\frac{\overline{R}(N_R,N_S,h,d)}{N_RN_S},
	\end{equation}
	which has units of ebits per second per satellite. Then, the goal is to take the satellite configuration that maximizes this figure of merit. In other words, our goal is to find
	\begin{equation}\label{eq-opt_sat_config}
		(N_R^{\star}(d),N_S^{\star}(d),h^{\star}(d))\coloneqq\argmax_{N_R,N_S,h} c(N_R,N_S,h,d)
	\end{equation}
	for any given distance $d$ between the two ground stations, where the optimization is constrained such that there is continuous coverage to the two ground stations for 24 hours and the transmission loss at any given time is less than 90 dB (see the Methods section for details). We suppress the dependence of the functions $N_R^{\star}$, $N_S^{\star}$, and $h^{\star}$ on the distance $d$ when it is understood from the context. We let
	\begin{equation}\label{eq-cost_func_normalized}
		C(h,d)\coloneqq \max_{N_R,N_S} c(N_R,N_S,h,d)
	\end{equation}
	be the figure of merit $c$ optimized over $N_R$ and $N_S$, with the constraint that both ground stations have continuous coverage over 24 hours and that the transmission loss at any time is less than 90~dB.
	
	\begin{table}
		\renewcommand{\arraystretch}{1.3}
		\centering
		\begin{tabular}{|c c c c c c|}
			\hline$d$ (km) & $h^{\star}$ (km) & $N_R^{\star}$ & $N_S^{\star}$ & $\overline{\eta}_{\text{dB}}$ & $\overline{R}$ (ebits/sec) \\ \hline\hline
			1500 & 1000 & 7 & 13 & 62.80 & 1321.32\\
			2500 & 1500 & 7 & 13 & 66.86 & 289.07\\
			3500 & 2000 & 8 & 10 & 72.93 & 70.02 \\
			4500 & 3000 & 8 & 9  & 77.64 & 20.52 \\ 
			5000 & 3500 & 8 & 9  & 79.75 & 12.03 \\\hline 
		\end{tabular}
		\caption{Satellite configurations $(N_R^{\star},N_S^{\star},h^{\star})$, as defined in Eq.~\eqref{eq-opt_sat_config}, corresponding to the maxima of the curves for the figure of merit $C(h,d)$ plotted in the central panel of Fig.~\ref{costfunc-graph1}. Also shown are the average loss $\overline{\eta}_{\text{dB}}\equiv\overline{\eta}_{\text{dB}}(N_R^{\star},N_S^{\star},h^{\star},d)$ and average rate $\overline{R}\equiv\overline{R}(N_R^{\star},N_S^{\star},h^{\star},d)$ over 24 hours for the optimal satellite configuration. See the Methods section for the formal definition of $\overline{\eta}_{\text{dB}}$ and $\overline{R}$.}\label{table_opt-params}
	\end{table}
	
	The results of our simulations are shown in Fig.~\ref{costfunc-graph1}. The complete set of results for all ground distances and satellite configurations considered is contained in the data files accompanying the paper. We first consider the quantity $N_{\text{opt}}(h,d)$ as a function of altitude $h$ for fixed ground-station separations $d$ (left panel of Fig.~\ref{costfunc-graph1}). In terms of the satellite configurations, we find that at higher altitudes more satellites per ring are required in general, while at lower altitudes generally more rings are required. In terms of the total number of satellites, we find that as the altitude increases the total number of satellites decreases. Interestingly, however, as we continue to increase the altitude we find that there are altitudes (between 5000~km and 6000~km) at which the total number of satellites reaches a minimum. Beyond this range of altitudes, the required number of satellites \textit{increases}. The presence of this minimum point gives us an indication of the altitudes at which satellites should be placed in order to minimize the total number of satellites. However, for these altitudes, the average entanglement-distribution rates are generally quite low, on the order of 10 ebits per second.

	Next, we consider the figure of merit $C(h,d)$ defined in Eq.~\eqref{eq-cost_func_normalized}. We plot this quantity for various values of the altitude $h$ and distance $d$ in the central panel of Fig.~\ref{costfunc-graph1}. In the right panel of Fig.~\ref{costfunc-graph1}, we plot the corresponding average entanglement-distribution rate over 24 hours. For all distances $d$, except for $d=500$~km, we find that there is an altitude $h$ at which $C(h,d)$ is maximal. These optimal altitudes, along with the values of $N_R$ and $N_S$ achieving the value of $C(h,d)$ and the corresponding average loss and average entanglement-distribution rate over 24 hours, are shown in Table~\ref{table_opt-params}. Given a desired distance between the ground stations, these optimal parameters can be used to decide on the number of satellites to put in the network and the altitude at which to put them so that there is continuous coverage, which then leads to particular values for the average loss and the average entanglement-distribution rate. Conversely, given a particular performance requirement (in terms of the entanglement-distribution rate), we can use our results to determine both the required satellite configuration and the required distance between the ground stations in order to achieve the desired rate. For example, using the plot on the right panel of Fig.~\ref{costfunc-graph1}, in order to achieve a rate greater than $10^3$ ebits per second on average in 24 hours, the satellite constellation altitude should be less than 2000~km (among the constellations considered), and the distance $d$ between the ground stations has to be roughly less than 1500~km.
	
	\begin{figure}
		\centering
		\includegraphics[width=0.9\columnwidth]{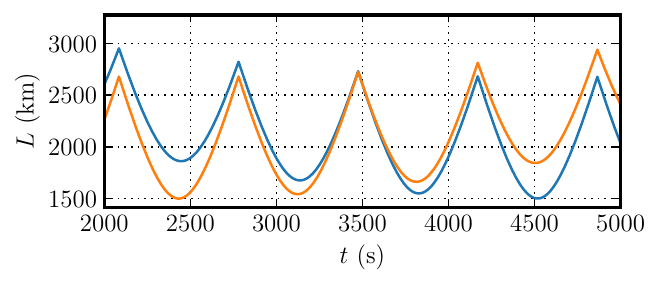}\\
		\includegraphics[width=0.9\columnwidth]{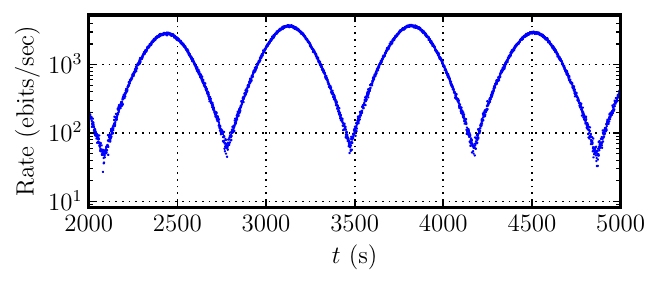}
		\caption{Entanglement distribution as a function of time to two ground stations at the equator. The ground stations are separated by $d=1000$~km with a satellite constellation given by $N_R=9$ satellite rings, $N_S=10$ satellites per ring, and altitude $h=1500$~km. We show a snapshot from 2000~s to 5000~s of our 24-hour simulation. (Top) The distance $L$ of each ground station to the satellite with the least total transmission loss. (Bottom) The corresponding entanglement-distribution rate as a function of time, assuming a source rate of $R_{\text{source}}=10^9$ ebits per second~\cite{CLZ+18}.}\label{fig-rate_vs_time}
	\end{figure}

	In Fig.~\ref{fig-rate_vs_time}, we plot the entanglement-distribution rate to two ground stations at the equator separated by a distance $d=1000$~km with a satellite constellation given by $N_R=9$ satellite rings, $N_S=10$ satellites per ring, and altitude $h=1500$~km. We also plot the distances of the ground stations to a satellite. We find that the rate exhibits a distinct oscillatory behavior with periodic bumps. In each bump, the rate increases as a satellite gets closer to the ground stations and decreases as the satellite passes by. All of the bumps in the rate have slightly different duration and slightly different peaks due to the fact that, at each time, the ground station pair is generally in view of multiple satellites, and we pick the satellite with the lowest transmission loss to the ground station pair (see the Methods section for further details). In general, therefore, each bump corresponds to a different satellite distributing entanglement to the two ground stations.
	
\begin{figure}
		\centering
		\includegraphics[width=\columnwidth]{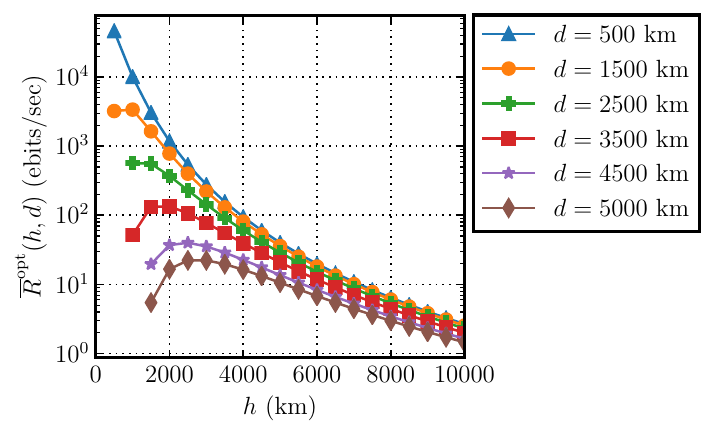}\\[0.2cm]
		\includegraphics[width=\columnwidth]{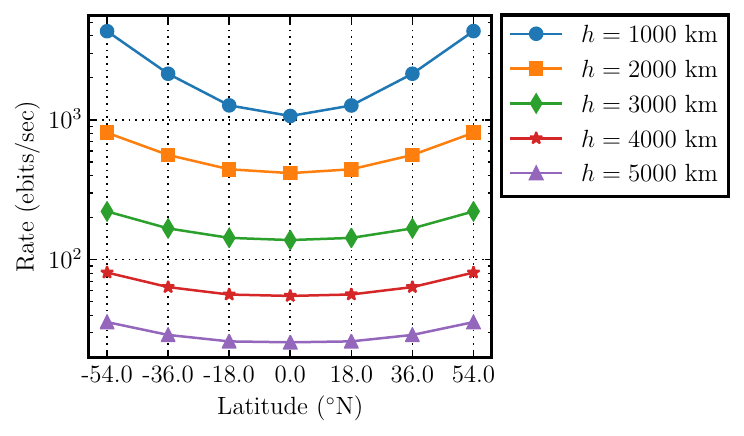}
		\caption{Average entanglement-distribution rates (over 24 hours) for two ground stations for various satellite constellations. In all cases, we assume that the satellites transmit entangled photon pairs at a rate of $R_{\text{source}}=10^9$ ebits per second \cite{CLZ+18}. (Top) Optimal rate (as defined in Eq.~\eqref{eq-opt_avg_ent_rate}) among all satellite configurations considered for two ground stations at the equator separated by a distance $d$. Each point in the plot corresponds to $N_R=20$ satellite rings and $N_S=20$ satellites per ring, because we find that this configuration achieves the maximum in Eq.~\eqref{eq-opt_avg_ent_rate}. (Bottom) Both ground stations at various latitudes. The ground stations are separated by approximately $18^{\circ}$ in longitude.  The satellite constellation consists of $N_R=15$ satellite rings with $N_S=15$ satellites per ring.}\label{fig-rates_two_ground_stations}
	\end{figure}

	Let us now consider optimal entanglement-distribution rates to the two ground stations, i.e., let us consider the quantity $\overline{R}^{\text{opt}}(h,d)$ defined in Eq.~\eqref{eq-opt_avg_ent_rate}. The results are shown in the top panel of Fig.~\ref{fig-rates_two_ground_stations}. We assume that the satellites transmit entangled photon pairs at a rate of $R_{\text{source}}=10^9$ ebits per second \cite{CLZ+18}. Unsurprisingly, for every pair $(h,d)$ of altitudes $h$ and distances $d$, the quantity $\overline{R}^{\text{opt}}(h,d)$ is attained by the satellite configuration that we considered that has the highest number of satellites, namely $N_R=20$ rings and $N_S=20$ satellites per ring. However, despite the sharp increase in the number of satellites, the rates are not much higher than those in the right panel of Fig.~\ref{costfunc-graph1}, which are obtained by optimizing our main figure of merit $C(h,d)$. The highest rate among all distances is around $4.6\times 10^4$ ebits per second, which is attained for a distance of $500$~km and altitude of $500$~km.

	\begin{figure*}
		\centering
		\includegraphics[width=\textwidth]{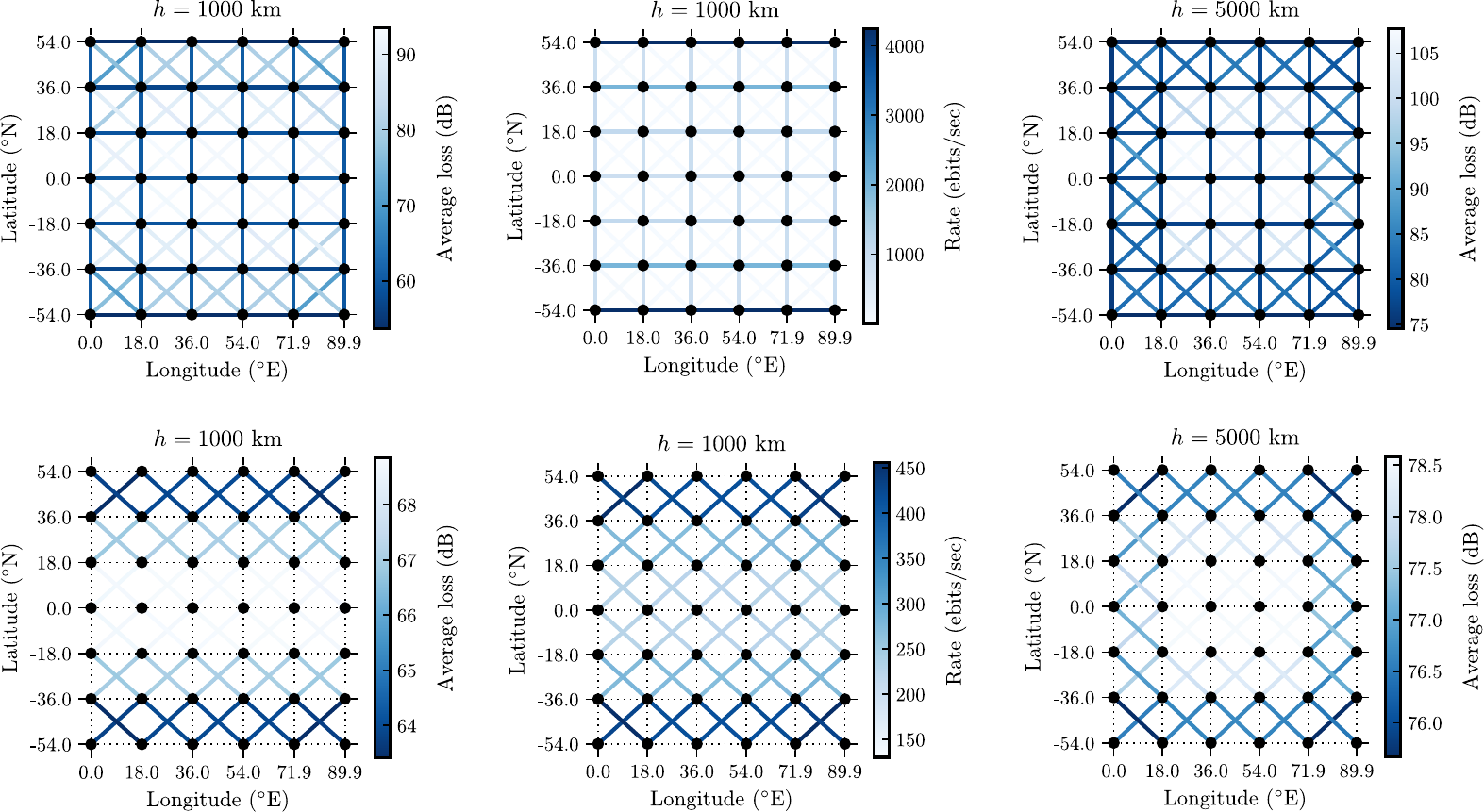}
		\caption{Average loss and rate (over 24 hours) for pairwise entanglement distribution for a collection of ground stations in a grid-like configuration. The nearest neighbors are separated by approximately $18^{\circ}$ in latitude and longitude. The satellite constellation consists of $N_R=15$ rings and $N_S=15$ satellites per ring, for a total of 225 satellites. Average rates in the central panel are calculated in a simple scenario without multimode transmission from the satellites and without multimode quantum memories at the ground stations. We assume that the satellites transmit entangled photon pairs at a rate of $R_{\text{source}}=10^9$ ebits per second \cite{CLZ+18}. (Top) Entanglement distribution to all possible nearest-neighbor pairs. (Bottom) Entanglement distribution only to diagonal nearest-neighbor pairs.}\label{fig-grid}
	\end{figure*}
	
	In the bottom panel of Fig.~\ref{fig-rates_two_ground_stations}, we display the results of our entanglement distribution simulations when both ground stations are at a different latitude, with $N_R=N_S=15$. Due to the fact that the satellites follow polar orbits in our network architecture, meaning that they congregate at the poles, the entanglement-distribution rates are higher for latitudes closer to the north and south poles than for the equator. This result also confirms that placing two ground stations at the equator is the worst-case scenario in terms of average loss (and thus average rate).

	Before continuing, let us remark that our technique for obtaining optimal satellite configurations for continuous global coverage, via optimization of the quantities defined in Eq.~\eqref{eq-cost_function} and Eq.~\eqref{eq-cost_func_normalized}, can be straightforwardly extended to an optimization procedure that consists of more than two ground stations; see the Methods section for details.

\subsection{Multiple ground stations}\label{sec-multiple_gr_stat}

	We now present the results of an entanglement distribution simulation consisting of multiple ground stations. We place 42 ground stations in a grid-like arrangement, with horizontal separation (i.e., separation in longitude) of approximately $18^{\circ}$ and vertical separation (i.e., separation in latitude) of approximately $18^{\circ}$. We use a satellite constellation of $N_R=15$ rings and $N_S=15$ satellites per ring, for a total of 225 satellites. In Fig.~\ref{fig-grid}, we display the average loss for nearest neighbor pairs over a simulation time of 24 hours.
	
	In the top plots of Fig.~\ref{fig-grid}, we consider all possible nearest-neighbor pairs in the simulation. As expected, the loss is lowest away from the equator (latitude $0^{\circ}$), because neighboring ground stations are closer to each other away from the equator, due to the curvature of the earth, and because of the nature of our satellite constellation (satellites congregate at the poles). We also find that diagonal nearest-neighbor pairs have higher losses compared to pairs that are horizontally or vertically separated. This can be explained by the fact that diagonally-separated ground stations are farther away from each other than horizontally- or vertically-separated ground-station pairs. Our strategy for assigning a satellite to a ground-station pair (see the Methods section) thus favors pairs that are horizontally or vertically separated. We also find that the maximum loss for a satellite altitude of $h=1000$~km is around 90 dB and the minimum loss is around 50 dB. For $h=5000$~km, the maximum loss is around 105 dB and the minimum loss is around 75~dB.
	
	In the bottom plots of Fig.~\ref{fig-grid}, we simulate a network such that the satellites can only distribute entanglement to diagonally-separated nearest-neighbor pairs. Now, since we do not allow entanglement distribution between horizontally- and vertically-separated pairs, we find that the maximum average loss decreases and the minimum average loss increases. We still find that ground-station pairs at latitudes farther away from the equator have lower loss.

	\begin{table*}
		\renewcommand{\arraystretch}{1.3}
		\centering
		\begin{tabular}{|c||c|c|c|c|c|c|c|}
			\hline \multirow{2}{*}{City pairs} & \multirow{2}{*}{\shortstack{Distance\\(km)}} & \multicolumn{6}{c|}{Average loss (dB)} \\ \cline{3-8}
			& & 500 km & 1000 km & 2000 km & 3000 km & 4000 km & 5000 km \\ \hline\hline
			Toronto -- New York City & 551 & 45.1 & 52.0 & 60.9 & 66.7 & 71.1 & 74.6 \\[0.1cm]
			Lijiang -- Delingha & 1200 & 50.6 & 52.9 & 60.5 & 66.3 & 70.7 & 74.3 \\[0.1cm]
			Houston -- Washington DC & 1922 & 75.1 & 66.9 & 73.7 & 78.3 & 81.1 & 83.1\\[0.1cm]
			Sydney -- Auckland & 2156 & 65.5 & 59.3 & 62.9 & 67.6 & 71.6 & 74.9 \\[0.1cm]
			New York City -- London & 5569 & $>90$ & $>90$ & 82.6 & 79.1 & 79.7 & 81.1 \\[0.1cm]
			Singapore -- Sydney & 6306 & $>90$ & $>90$ & $>90$ & 83.3 & 82.5 & 83.2 \\[0.1cm]
			London -- Mumbai & 7191 & $>90$ & $>90$ & $>90$ & $>90$ & 89.0 & 88.3 \\ \hline
		\end{tabular}
		\caption{Average loss over a 24-hour period between select pairs of major global cities for a constellation of 400 satellites ($N_R=N_S=20$) at various altitudes. The following cities are included in the simulation: Toronto, New York City, London, Singapore, Sydney, Auckland, Rio de Janeiro, Baton Rouge, Mumbai, Johannesburg, Washington DC, Lijiang, Ngari, Delingha, Nanshan, Xinglong, and Houston.}\label{table-hubs_losses}
	\end{table*}
	
	In the central panels of Fig.~\ref{fig-grid}, we plot average entanglement-distribution rates in a simple scenario without multimode transmission from the satellites and without multimode quantum memories at the ground stations. We assume that the satellites transmit entangled photon pairs at a rate of $R_{\text{source}}=10^9$ ebits per second \cite{CLZ+18}. In the case of entanglement distribution to all nearest-neighbor pairs (top part of the central panel of Fig.~\ref{fig-grid}), the maximum average rate is around 4000 ebits per second, and this occurs for horizontally separated ground stations at latitudes of $54^{\circ}$N and $-54^{\circ}$N. For entanglement distribution only to diagonally-separated nearest-neighbor pairs (bottom part of the central panel of Fig.~\ref{fig-grid}), the maximum average rate is around 450 ebits per second. It is possible to compensate for the loss by having multimode signal transmission from the satellites and by including multimode quantum memories at the ground stations, which would increase the average rates.

\subsection{Entanglement distribution between major global cities}
	
	Although the ultimate goal of a satellite-based quantum internet is to have satellites distribute entanglement between any collection of nodes on the ground, an example of which we considered above, satellite-based quantum communication networks will likely have a hybrid form in the near term. In a hybrid network, the satellites distribute entanglement to major global cities, which act as hubs that then distribute entanglement to smaller nearby cities using ground-based links (see Fig.~\ref{fig-hybrid_network}). With this in mind, we now consider entanglement distribution between pairs of major global cities. We run a 24-hour simulation with a satellite constellation of 400 satellites, with $N_R=N_S=20$, at altitudes of $h=500~\text{km}$, $1000~\text{km}$, $2000~\text{km}$, $3000~\text{km}$, $4000~\text{km}$, and $5000~\text{km}$. We include the following cities in the simulation: Toronto, New York City, London, Singapore, Sydney, Auckland, Rio de Janeiro, Baton Rouge, Mumbai, Johannesburg, Washington DC, Lijiang, Ngari, Delingha, Nanshan, Xinglong, and Houston. The Lijiang-Delingha pair is chosen for comparison to a recent experiment \cite{YCL17}. The simulation results are shown in Table~\ref{table-hubs_losses}.
	
	From Table~\ref{table-hubs_losses}, we see that at around a distance of 6300~km, which is the distance between Singapore and Sydney, we can only obtain an average loss less than 90 dB for altitudes greater than 2000~km. Similarly, entanglement distribution between London and Mumbai (which are 7200~km apart) at an average loss less than 90 dB is possible only for an altitude greater than 3000~km. These results suggest that, using our constellation of 400 satellites, a distance of around 7500~km is the highest for which entanglement distribution at a loss less than 90 dB can be achieved. Indeed, for Houston and London (which are 7800~km apart), we find that the average loss is greater than 90 dB for all of the satellite altitudes that we consider.

\subsection{Comparison to ground-based entanglement distribution}\label{sec-sats_vs_repeaters}

	Let us now compare the entanglement-distribution rates obtained with satellites to the rates that can be obtained via ground-based photon transmission through optical fiber with the assistance of quantum repeaters. In particular, we compare the rates in the top panel of Fig.~\ref{fig-rates_two_ground_stations} for two ground stations at the equator separated by a distance $d$ between 100~km and 2000~km to ground-based repeater chains with endpoints the same distance $d$ apart. For the latter, we suppose that the distance $d$ between the endpoints is split into $M$ elementary links by $(M-1)$ equally-spaced quantum repeaters. We place a source at the center of each elementary link that transmits entangled photon pairs to the nodes at the ends of the elementary link. We assume that the probability of establishing an elementary link is $p=\e^{-\alpha\frac{d}{M}}$, where $\alpha=\frac{1}{22\text{ km}}$ \cite{DKD18}, and we also assume that all repeater nodes are equipped with $N_{\text{mem}}$ quantum memories facing each of its nearest neighbors. Under these conditions, the rate $R_{M,N_{\text{mem}}}$ (in ebits per second) of entanglement distribution between the endpoints is
	\begin{equation}\label{eq-rate_upper_bound}
		R_{M,N_{\text{mem}}}=\frac{cN_{\text{mem}}}{2(d/M)}\frac{1}{W_{M,N_{\text{mem}}}},
	\end{equation}
	where $c$ is the speed of light and
	\begin{equation}
		W_{M,N_{\text{mem}}}=\sum_{n=1}^{\infty}\left(1-\left(1-(1-p)^{n-1}\right)^M\right)^{N_{\text{mem}}}.
	\end{equation}
	(See the Methods section for details.) Note that our assumption that $p=\e^{-\alpha\frac{d}{M}}$ is the best-case scenario in which the sources fire perfect Bell pairs (so that no entanglement purification is required) and the Bell measurements for entanglement swapping are deterministic. Furthermore, the formula in Eq.~\eqref{eq-rate_upper_bound} holds in the case that the quantum repeaters have perfect read-write efficiency and have infinite coherence time.
	
\begin{figure}
		\centering
		\includegraphics[width=0.85\columnwidth]{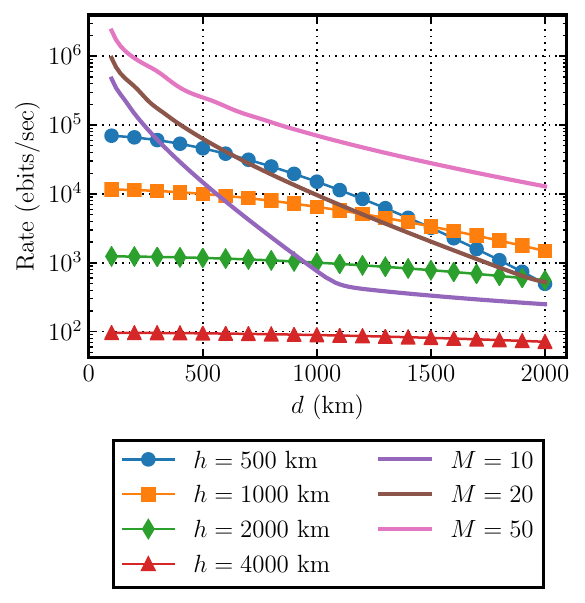}
		\caption{Comparison of satellite-based entanglement distribution to ground-based repeater-assisted entanglement distribution. We consider two ground stations at the equator separated by a distance $d$, with $N_R=20$ satellite rings and $N_S=20$ satellites per ring. We compare to a ground-based repeater chain of the same distance $d$ consisting of $M$ elementary links of equal length and $N_{\text{mem}}=50$ quantum memories per elementary link. The rate is given by Eq.~\eqref{eq-rate_upper_bound}.}\label{fig-rates_repeaters_vs_sats}
	\end{figure}

	In Fig.~\ref{fig-rates_repeaters_vs_sats}, we compare the rate in Eq.~\eqref{eq-rate_upper_bound} with $N_{\text{mem}}=50$ to the rates shown in the top panel of Fig.~\ref{fig-rates_two_ground_stations}. For an altitude of 500~km, we find that the quantum repeater scheme with $M=50$ elementary links outperforms the satellite-based scheme for all distances up to 2000~km. However, for $M=10$ and $M=20$ elementary links, we find that there are critical distances beyond which satellites can outperform the ground-based repeater schemes. For example, for an altitude of 500~km, the satellite-based scheme outperforms the $M=20$ quantum repeater scheme beyond approximately 600~km and the $M=10$ scheme beyond approximately 300~km. For an altitude of 1000~km, the satellite-based scheme outperforms the $M=20$ repeater scheme beyond approximately 1200~km. Similarly, for an altitude of 2000~km, the satellite-based scheme outperforms the quantum repeater scheme beyond approximately 900~km. For an altitude of 4000~km, the satellite-based rates are lower than the quantum repeater rates for all values of $M$ considered.
	
	Currently, satellite-based schemes are arguably more viable, because high-coherence-time quantum memories (which are not widely available) are not required. However, the monetary cost of the satellites, along with other overhead monetary costs (e.g., launch costs), can make implementing a satellite-based entanglement-distribution network challenging. Furthermore, local weather conditions and background photons during the daytime make it difficult to achieve the continuous coverage assumed here, which ultimately results in lower entanglement-distribution rates. On the other hand, ground-based quantum repeater schemes can achieve higher rates than satellite-based schemes, but this occurs only when the number of repeater nodes is quite high, the number of quantum memories per repeater node is high, and the coherence times of the memories is high. In addition, quantum memories currently exist mostly in a laboratory environment and are not at the stage of development that they can be widely deployed in the field, and they certainly do not have high enough coherence times to achieve the rates presented here.

\section{Discussion and conclusions}

	In this paper, we explored the possibility of using satellites for a global-scale quantum communications network. Our network architecture consists of a constellation of satellites in polar orbits around the earth that transmit entangled photon pairs to ground stations (see Fig.~\ref{fig-sat_architecture}). By defining a figure of merit that takes into account both the number of satellites as well as satellite-to-ground entanglement-distribution rates, we provided estimates on the number of satellites needed to maintain full 24-hour coverage at a high rate based on the maximum value of the figure of merit. Using our figure of merit to decide the number of satellites in the network, we estimated the transmission loss and entanglement-distribution rates that can be achieved for two ground stations placed at various latitudes, for multiple ground stations at various locations in a grid-like arrangement, and for multiple major global cities in a hybrid satellite- and ground-based network in which the cities can act as hubs that receive entanglement from satellites and disperse it to surrounding locations via ground-based links. Finally, we compared the achievable entanglement-distribution rates for two ground stations using satellites to achievable entanglement-distribution rates using ground-based links with quantum repeaters. With a large enough number of repeater nodes, along with a high enough number of high-coherence-time quantum memories at each node, it is possible to obtain entanglement-distribution rates that surpass those obtained with satellites. However, satellite-based schemes operating without quantum repeaters can, in certain cases, outperform quantum repeater schemes, with drawbacks being that a relatively high number of satellites is required and that adverse weather conditions can prevent continuous operations and thus reduce the rate. These drawbacks appear to be less prohibitive in the near term than the major drawback of ground-based, repeater-assisted entanglement distribution, which is that quantum memories with very high coherence times are simply not widely available. Therefore, it appears that a satellite-based scheme will remain the preferred option over ground-based repeater schemes into the near term, especially with the improving miniaturization and increasing fidelity of entanglement sources \cite{Kang:16,TCT+16} and the decreasing cost and miniaturization of satellites \cite{JH13,Cubesat2017,Nanobob2018}.
	
	Our analysis of a global, satellite-based quantum internet opens the door to plenty of further study. For example, our simulations can be refined by taking into account local weather conditions. Our optimization procedure can also be extended to include more than two ground stations (see the Methods section). It would also be interesting to compare other types of satellite constellations, much like those studied in Refs.~\cite{LA98,LFP98}. Finally, to have a genuine quantum network requires efficient routing algorithms. It would be interesting to explore entanglement routing in a satellite network along the lines of, e.g., Refs.~\cite{GPA99,LLKK00,Handley18} in the classical setting.
	
	In summary, the broad-scope vision is to have a quantum-connected world, similar to today's internet, where users across the globe can share quantum information for any desirable task. In our view, the backbone of such a network is built on local and global quantum entanglement, in which intercontinentally-separated ground stations located in major cities act as entanglement hubs connecting the local network users of one city to those of another (Fig. \ref{fig-hybrid_network}). Hybrid networks interfacing space-based quantum communication platforms with ground-based quantum repeaters will make this vision a real possibility.

\section{Methods}

\subsection{Loss model}

	In the absence of spurious background photons, the transmission of photons from satellites to ground stations is modeled well by a bosonic pure-loss channel with transmittance $\eta_{\text{sg}}$ \cite{Serafini_book}. For single-photon polarization qubits (with a dual-rail encoding), transmission through the pure-loss channel corresponds to an erasure channel \cite{BH14}. That is, given a single-photon polarization density matrix $\rho$, the evolution of $\rho$ is given as
	\begin{equation}\label{eq-erasure_channel}
		\rho\mapsto\eta_{\text{sg}}\rho +(1-\eta_{\text{sg}})\dyad{\text{vac}}
	\end{equation} 
	where $\dyad{\text{vac}}$ is the vacuum state. Hence, with probability $\eta_{\text{sg}}$, the dual-rail qubit is successfully transmitted and with probability $1-\eta_{\text{sg}}$ the qubit is lost. For the transmission of a pair of single-photon dual-rail qubits, let $\eta^{(1)}_{\text{sg}}$ and $\eta^{(2)}_{\text{sg}}$ be the transmittances of the two pure-loss channels. Then, with probability $\eta^{(1)}_{\text{sg}}\eta^{(2)}_{\text{sg}}$, both qubits are successfully transmitted and with probability $1-\eta_{\text{sg}}^{(1)}\eta_{\text{sg}}^{(2)}$ at least one of the qubits is lost \cite{DKD18}. In the following subsection, we consider photon transmission in the presence of background photons.

	The transmittance $\eta_{\text{sg}}$ generally depends on atmospheric conditions (such as turbulence and weather conditions) and on orbital parameters (such as altitude and zenith angle) \cite{Vogel2017weather,Vogel2019atmlinks,LKB+19}. In general, we can decompose $\eta_{\text{sg}}$ as 
	\begin{equation}
		\eta_{\text{sg}}=\eta_{\text{fs}}\eta_{\text{atm}}\label{eq-eta_sg}
	\end{equation}
	where $\eta_{\text{fs}}$ is the free-space transmittance and $\eta_{\text{atm}}$ is the atmospheric transmittance. Free-space loss occurs due to diffraction (i.e., beam broadening) over the channel and due to the use of finite-sized apertures at the receiving end. These effects cause $\eta_{\text{fs}}$ to scale as the inverse-distance squared in the far-field regime. Atmospheric loss occurs due to absorption and scattering in the atmosphere and scales exponentially with distance as a result of the Beer-Lambert law \cite{BH08,Andrews2005randomedia,KJK_book}. However, since atmospheric absorption is relevant only in a layer of thickness 10-20~km above the earth's surface \cite{KJK_book}, free-space diffraction is the main source of loss in space-based quantum communication. In order to characterize the free-space and atmospheric transmittances with simple analytic expressions, we ignore turbulence-induced effects in the lower atmosphere, such as beam profile distortion, beam broadening (prominent for uplink communication \cite{KJK_book,BMH+13}), and beam wandering (see, e.g., Ref. \cite{Vogel2019atmlinks}). Note that turbulence effects can be corrected using classical adaptive optics \cite{KJK_book}. We also ignore the inhomogeneous density profile of the atmosphere, which can lead to path elongation effects at large zenith angles. A comprehensive analysis of loss can be found in Refs.~\cite{Andrews2005randomedia,Vogel2019atmlinks}.

	\begin{table}
		\renewcommand{\arraystretch}{1.3}
		\centering
		\begin{tabular}{|>{\centering\arraybackslash}m{1.3cm}  >{\centering\arraybackslash}m{3.5cm}  >{\centering\arraybackslash}m{1.4cm} |}
			\hline
			Parameter & Definition & Value \\\hline\hline
			$r$ & Receiving aperture radius & 0.75 m \\[0.2cm] 
			$w_0$ & Initial beam waist & 2.5 cm \\[0.2cm]
			$\lambda$ & Wavelength of satellite-to-ground signals & 810 nm \\[0.6cm]
			$\eta_{\text{atm}}^{\text{zen}}$ & Atmospheric transmittance at zenith & 0.5 at 810 nm \cite{BMH+13} \\ \hline
		\end{tabular}
		\caption{Parameters used in the modeling of loss from satellites to ground stations.}\label{table-parameters}
	\end{table}

	Consider the lowest-order Gaussian spatial mode for an optical beam traveling a distance $L$ between the sender and receiver, with a circular receiving aperture of radius $r$. Then, the free-space transmittance $\eta_{\text{fs}}$ is given by \cite{SveltoBook}
	\begin{equation}\label{eq-fs_transmittance}
		\eta_{\text{fs}}(L)=1-\exp\left(-\frac{2r^2}{w(L)^2}\right)  .  
	\end{equation}  
where
	\begin{equation}
		w(L)\coloneqq w_{0}\sqrt{1+\left(\frac{L}{L_{R}}\right)^2}
	\end{equation}
	is the beam waist at a distance $L$ from the focal region ($L=0$), $L_{R}\coloneqq\pi w_{0}^2\lambda^{-1}$ is the Rayleigh range, $\lambda$ is the wavelength of the optical mode, and $w_0$ is the initial beam-waist radius.

	We model the atmosphere as a homogeneous absorptive layer of finite thickness in order to characterize $\eta_{\text{atm}}$. Uniformity of the atmospheric layer then implies uniform absorption (at a given wavelength), such that $\eta_{\text{atm}}$ depends only on the optical path traversed through the atmosphere. Under these assumptions, and using the Beer-Lambert law \cite{BH08}, for small zenith angles we have that
	\begin{equation}\label{eq-atmospheric_transmittance_zenith}
		\eta_{\text{atm}}(L,h)=\left\{\begin{array}{l l} (\eta_{\text{atm}}^{\text{zen}})^{\sec\zeta}, & \text{if } -\frac{\pi}{2}<\zeta<\frac{\pi}{2},\\[0.2cm] 0, & \text{if } |\zeta|\geq\frac{\pi}{2}, \end{array}\right.
	\end{equation}
	with $\eta_{\text{atm}}^{\text{zen}}$ the transmittance at zenith ($\zeta=0$). For $|\zeta|>\frac{\pi}{2}$, we set $\eta_{\text{atm}}=0$, because the satellite is over the horizon and thus out of sight. The zenith angle $\zeta$ is given by
	\begin{equation}\label{eq-sec_zen}
		\cos\zeta=\frac{h}{L}-\frac{1}{2}\frac{L^2-h^2}{R_{E}L}
	\end{equation}
	for a circular orbit of altitude $h$, with $R_E\approx 6378$~km being the earth's radius.
	
	Note that the model of atmospheric transmittance given by Eq.~\eqref{eq-atmospheric_transmittance_zenith} and Eq.~\eqref{eq-sec_zen} is quite accurate for small zenith angles \cite{KJK_book}. However, for space-based quantum communication at or near the horizon (i.e., for $\zeta=\pm\pi/2$), more exact methods relying on the standard atmospheric model must be used \cite{Vogel2019atmlinks}. In practice, it makes sense to set $\eta_{\text{atm}}=0$ at large zenith angles, effectively severing the quantum channel, because the loss will typically be too high for the link to be practically useful.
	
	\begin{figure}
		\centering
		\includegraphics[width=\columnwidth]{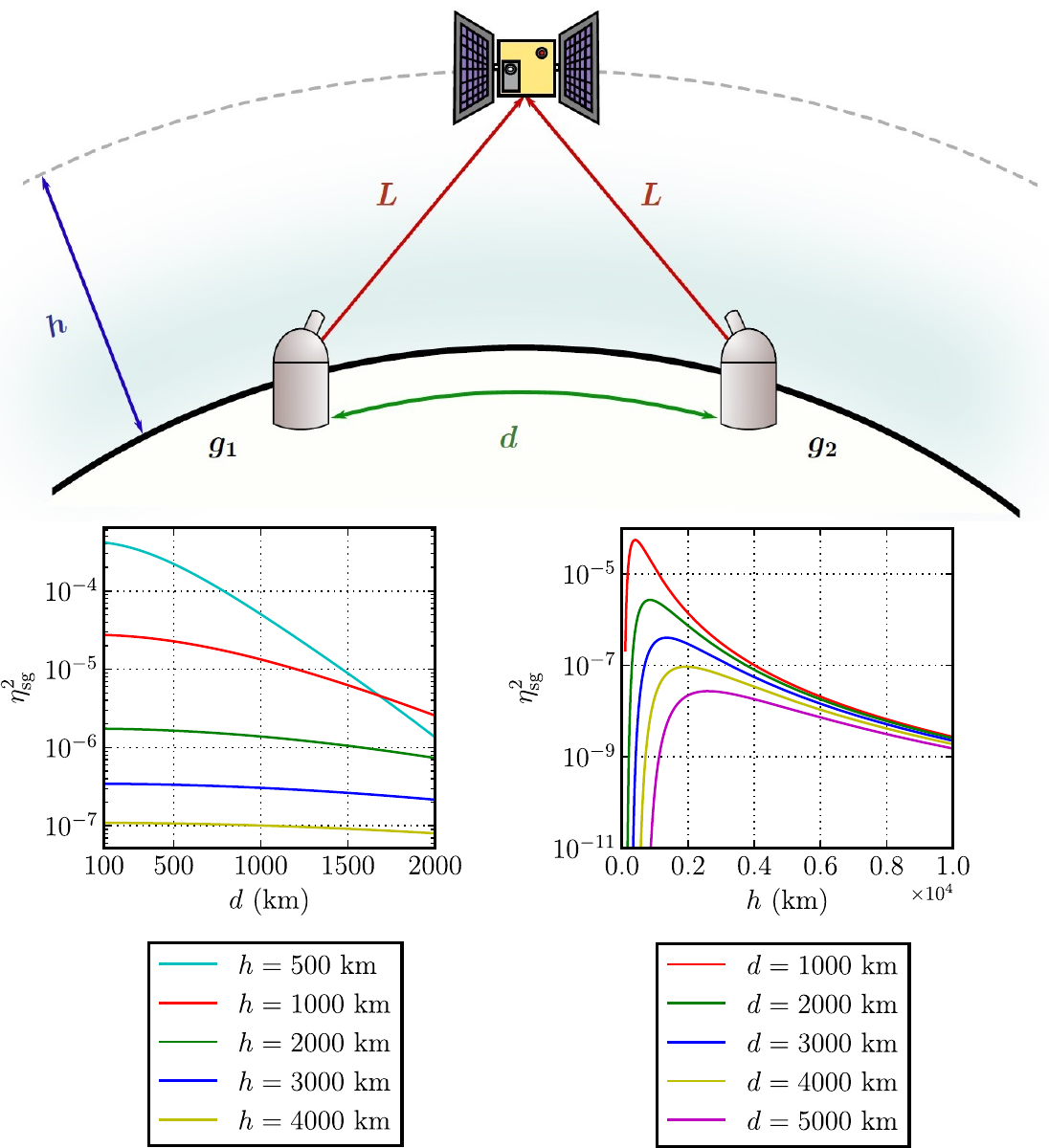}
		\caption{Optical satellite-to-ground transmission. The total transmittance is given by $\eta_{\text{sg}}=\eta_{\text{fs}}\eta_{\text{atm}}$, where the free-space transmittance $\eta_{\text{fs}}$ given by Eq.~\eqref{eq-fs_transmittance}, and the atmospheric transmittance $\eta_{\text{atm}}$ is given by Eq.~\eqref{eq-atmospheric_transmittance_zenith}. (Top) Two ground stations $g_1$ and $g_2$ are separated by a distance $d$ with a satellite at altitude $h$ at the midpoint. Both ground stations are the same distance $L$ away from the satellite, so that the total transmittance for two-qubit entanglement transmission (one qubit to each ground station) is $\eta_{\text{sg}}^2$. (Bottom) Plots of the transmittance $\eta_{\text{sg}}^2$ as a function of $d$ and $h$.}\label{fig-atmosphere_geometry}
	\end{figure}
	
	To summarize, the following parameters characterize the total loss $\eta_{\text{sg}}=\eta_{\text{fs}}\eta_{\text{atm}}$: the initial beam waist $w_0$, the receiving aperture radius $r$, the wavelength $\lambda$ of the satellite-to-ground signals, and the atmospheric transmittance $\eta_{\text{atm}}^{\text{zen}}$ at zenith. See Table~\ref{table-parameters} for the values that we take for these parameters in our simulations.
	
	Using the values in Table~\ref{table-parameters}, we plot in Fig.~\ref{fig-atmosphere_geometry} (bottom) the total transmittance as a function of the ground distance $d$ between two ground stations with a satellite at the midpoint; see Fig.~\ref{fig-atmosphere_geometry} (top). We observe that for larger ground separations the total transmittance $\eta_{\text{sg}}^2$ is actually larger for a higher altitude than for a lower altitude; for example, beyond approximately $d=1600$~km the transmittance for $h=1000$~km is larger than for $h=500$~km. We also observe that there are altitudes at which the transmittance is maximal. Intuitively, beyond the maximum point, the atmospheric contribution to the loss is less dominant, while below the maximum (i.e., for lower altitudes) the atmosphere is the dominant source of loss. This feature is unique for optical transmission from satellite to ground.

\subsection{Noise model}
	
	We now consider photon transmission in the presence of background photons. We analyze the scenario in which a source generates an entangled photon pair and distributes the individual photons to two parties, Alice ($A$) and Bob ($B$). We allow the distributed photons to mix with spurious photons (noise) from an uncorrelated thermal source, assuming a low thermal background (which can be ensured via stringent filtering). We then determine, in the high loss and low noise regime, the fidelity of the distributed entangled photon pair.
	
	First, consider a tensor product of thermal states for the horizontal and vertical polarization modes:
	\begin{align}
    	\Theta^{\Bar{n}_H}\otimes\Theta^{\Bar{n}_V}&=\Bigg(\sum_{n=0}^\infty\left(\frac{\Bar{n}_H^n}{(\Bar{n}_H+1)^{n+1}}\right) \dyad{n}\Bigg)\nonumber\\
    	&\qquad\otimes\left(\sum_{n=0}^\infty\left(\frac{\Bar{n}_V^n}{(\Bar{n}_V+1)^{n+1}}\right)\dyad{n}\right),
	\end{align}
	where $\Bar{n}_k$ is the average number of photons in the thermal state for the polarization mode $k$. We assume this state comes from an incoherent source with no polarization preference (e.g., the sun), such that $\Bar{n}_H=\Bar{n}_V\eqqcolon\Bar{n}/2$. Furthermore, we assume some (non-polarization) filtering procedure, which reduces the number of background thermal photons, such that $\Bar{n}\ll1$. We then rewrite the above state to first order in the small parameter $\Bar{n}$:
	\begin{align}
    	\Theta^{\frac{\Bar{n}}{2}}\otimes\Theta^{\frac{\Bar{n}}{2}}&\approx\left(\left(1-\frac{\Bar{n}}{2}\right)\dyad{0} +\frac{\Bar{n}}{2}\dyad{1}\right)\nonumber\\
    	&\qquad\otimes\left(\left(1-\frac{\Bar{n}}{2}\right)\dyad{0} +\frac{\Bar{n}}{2}\dyad{1}\right) \\
    	&\approx(1-\Bar{n})\dyad{\text{vac}}+\frac{\Bar{n}}{2}\left(\dyad{H}+\dyad{V}\right),\label{eq-therm_expand}
	\end{align}
	where $\ket{\text{vac}}=\ket{0}\otimes\ket{0}$, and
	\begin{align}
    	\ket{H}&\coloneqq\ket{1}\otimes\ket{0},\\
    	\ket{V}&\coloneqq\ket{0}\otimes\ket{1}.
	\end{align}
	We thus define our approximate thermal background state as
	\begin{equation}\label{eq-approx_th_state}
		\widetilde{\Theta}^{\Bar{n}}\coloneqq(1-\Bar{n})\dyad{\text{vac}}+\frac{\Bar{n}}{2}\left(\dyad{H}+\dyad{V}\right),
	\end{equation}
	which serves as a good approximation to a low thermal background. The transmission channel from the source to the ground is then approximately
	\begin{widetext}
	\begin{equation}\label{eq-noisy_transmission_channel}
		\mathcal{L}_{\eta_{\text{sg}},\Bar{n}}(\rho_{A_1A_2})\coloneqq\Tr_{E_1E_2}[(U_{A_1E_1}^{\eta_{\text{sg}}}\otimes U^{\eta_{\text{sg}}}_{A_2E_2})(\rho_{A_1A_2}\otimes\widetilde{\Theta}_{E_1E_2}^{\Bar{n}})(U^{\eta_{\text{sg}}}_{A_1E_1}\otimes U^{\eta_{\text{sg}}}_{A_2E_2})^\dagger],
	\end{equation}
	\end{widetext}
	where $U^{\eta_{\text{sg}}}$ is the beamsplitter unitary (see, e.g., Ref.~\cite{Serafini_book}), and $A_1$ and $A_2$ refer to the horizontal and vertical polarization modes, respectively, of the dual-rail quantum system being transmitted; similarly for $E_1$ and $E_2$. Note that for $\Bar{n}=0$, the transformation given by Eq.~\eqref{eq-noisy_transmission_channel} is equal to the transformation in \eqref{eq-erasure_channel}. For a source state $\rho_{AB}^S$, with $A\equiv A_1A_2$ and $B\equiv B_1B_2$, the quantum state shared by Alice and Bob after transmission of the state $\rho_{AB}^S$ from the satellite to the ground stations is
	\begin{equation}\label{eq-transmission_noisy_output}
		(\mathcal{L}_{\eta_{\text{sg}}^{(1)},\Bar{n}_1}\otimes\mathcal{L}_{\eta_{\text{sg}}^{(2)},\Bar{n}_2})(\rho_{AB}^S).
	\end{equation}

	Let us first assume that we have an ideal two-photon source, which generates one of the four two-photon polarization-entangled Bell states, i.e., a state of the form $\rho^S=\Phi^{\pm}\coloneqq\dyad{\Phi^{\pm}}$ or $\rho^S=\Psi^{\pm}\coloneqq\dyad{\Psi^{\pm}}$, where
	\begin{align}
		\ket{\Phi^{\pm}}&\coloneqq\frac{1}{\sqrt{2}}(\ket{H,H}\pm\ket{V,V}),\\
		\ket{\Psi^{\pm}}&\coloneqq\frac{1}{\sqrt{2}}(\ket{H,V}\pm\ket{V,H}).
	\end{align}
	After transmission, we assume post-selection on coincident events, along with high loss and low noise ($\eta_{\text{sg}}^{(1)},\eta_{\text{sg}}^{(2)},\Bar{n}\ll 1$). The post-selection allows one to discard any occurrence in which one site registers a photon and the other does not. Furthermore, under the high-loss and low-noise assumptions, we can discard potential four-photon and three-photon occurrences, as these occur with negligible probability compared to the two-photon events. We thus focus our full attention on the two-photon state corresponding to one photon received at Alice's site and one photon received at Bob's site. Mathematically, this (unnormalized) state is given by
	\begin{equation}
		\Pi_{AB}(\mathcal{L}_{\eta_{\text{sg}}^{(1)},\Bar{n}_1}\otimes\mathcal{L}_{\eta_{\text{sg}}^{(2)},\Bar{n}_2})(\rho_{AB}^S)\Pi_{AB},
	\end{equation}
	where
	\begin{multline}
		\Pi_{AB}\coloneqq(\dyad{H}_A+\dyad{V}_A)\\\otimes(\dyad{H}_B+\dyad{V}_B)
	\end{multline}
	is the projection onto the two-photon-coincidence subspace. With $\rho_{AB}^S=\Phi_{AB}^{\pm}$, it is straightforward to show that
	\begin{align}
		&\Pi_{AB}(\mathcal{L}_{\eta_{\text{sg}}^{(1)},\Bar{n}_1}\otimes\mathcal{L}_{\eta_{\text{sg}}^{(2)},\Bar{n}_2})(\Phi_{AB}^{\pm})\Pi_{AB}\nonumber\\
		&\quad=\frac{1}{2}(x_1x_2+y_1y_2\pm z_1z_2)\Phi_{AB}^+\nonumber\\
		&\qquad+\frac{1}{2}(x_1x_2+y_1y_2\mp z_1z_2)\Phi_{AB}^-\nonumber\\
		&\qquad+\frac{1}{2}(x_1y_2+y_1x_2)\Psi_{AB}^+\nonumber\\
		&\qquad+\frac{1}{2}(x_1y_2+y_1x_2)\Psi_{AB}^-,\label{eq-final_state}
	\end{align}
	where
	\begin{align}
		x_{1}&\coloneqq (1-\Bar{n}_{1})\eta_{\text{sg}}^{(1)}+\frac{\Bar{n}_{1}}{2}((1-2\eta_{\text{sg}}^{(1)})^2+(\eta_{\text{sg}}^{(1)})^2),\\
		y_{1}&\coloneqq \frac{\Bar{n}_{1}}{2}(1-\eta_{\text{sg}}^{(1)})^2,\\
		z_{1}&\coloneqq (1-\Bar{n}_{1})\eta_{\text{sg}}^{(1)}-\Bar{n}_{1}\eta_{\text{sg}}^{(1)}(1-2\eta_{\text{sg}}^{(1)}),
	\end{align}
	with analogous definitions for $x_2,y_2,z_2$. The fidelity of this quantum state conditioned on one photon received by Alice and one photon received by Bob is therefore
	\begin{align}
		F_{\Phi^{\pm}}&\coloneqq \frac{\bra{\Phi^+}\Pi(\mathcal{L}_{\eta_{\text{sg}}^{(1)},\Bar{n}_1}\otimes\mathcal{L}_{\eta_{\text{sg}}^{(2)},\Bar{n}_2})(\Phi_{AB}^{\pm})\Pi\ket{\Phi^+}}{\Tr[\Pi(\mathcal{L}_{\eta_{\text{sg}}^{(1)},\Bar{n}_1}\otimes\mathcal{L}_{\eta_{\text{sg}}^{(2)},\Bar{n}_2})(\Phi_{AB}^{\pm})\Pi]}\\
		&=\frac{\frac{1}{2}(x_1x_2+y_1y_2\pm z_1z_2)}{(x_1+y_1)(x_2+y_2)}.
	\end{align}
	Assuming that $\eta_{\text{sg}}^{(1)}=\eta_{\text{sg}}^{(2)}=\eta_{\text{sg}}$ and $\Bar{n}_1=\Bar{n}_2=\Bar{n}$, so that $x_1=x_2$, $y_1=y_2$, and $z_1=z_2$, and under the high-loss and low-noise assumption, for $\rho^S=\Phi^+$ this reduces to
	\begin{equation}\label{eq-fidelity_ideal}
		F_{\Phi^+}\approx \frac{1}{4}\left(1+\frac{3}{\left(1+\frac{\Bar{n}}{\eta_{\text{sg}}}\right)^2}\right).
	\end{equation}
	The ratio $\frac{\eta_{\text{sg}}}{\Bar{n}}$ is just the local signal-to-noise ratio (SNR). Thus, assuming a fidelity constraint $F\gtrsim F^\star$, we obtain the following bound on the SNR needed at each site in order to maintain a fidelity of $F^\star$ during operation:
	\begin{equation}
		\text{SNR}\coloneqq\frac{\eta_{\text{sg}}}{\Bar{n}}\gtrsim\frac{1}{\left(\sqrt{\frac{3}{4F^\star-1}}-1\right)}\approx\frac{3}{2}(1-F^\star)^{-1},
	\end{equation}
	Here, we have assumed that the fidelity lies within some small range close to one (e.g., $.95\leq F^\star\leq1$) and expanded to first order in $1-F^\star$. As an example, consider $F^\star=.99$. Then, we must have $\text{SNR}\gtrsim 150$ at each site. Given that $\eta_{\text{sg}}\sim 10^{-3}$, this implies a constraint on the number of background photons per detection window of $\Bar{n}\lesssim 7\times 10^{-6}$.

\subsubsection{Non-ideal Bell states}

\begin{figure}
    	\centering
    	\includegraphics[width=0.85\columnwidth]{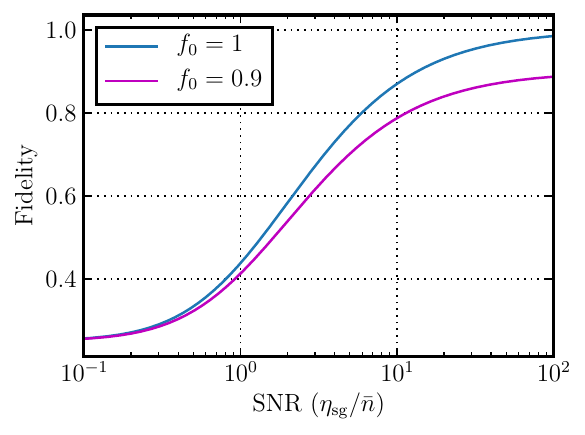}
    	\caption{Fidelity of satellite-to-ground entanglement transmission as a function of the signal-to-noise ratio (SNR) of the transmission medium. The source state is in Eq.~\eqref{eq-souce_state_imperfect}, and the fidelity after transmission is given by Eq.~\eqref{eq-fidelity_nonideal_bell_state}.} \label{fig-fidelity_snr}
	\end{figure}

	Let us now consider an initially imperfect Bell state generated by a non-ideal entangled photon-pair source. Specifically, we consider the state
	\begin{equation}\label{eq-souce_state_imperfect}
		\rho^S(f_0)\coloneqq f_0\Phi^++\left(\frac{1-f_0}{3}\right)(\Phi^-+\Psi^++\Psi^-),
	\end{equation}
	where $f_0$ is the initial fidelity. Using the fact that
	\begin{align}
		&\Pi_{AB}(\mathcal{L}_{\eta_{\text{sg}}^{(1)},\Bar{n}_1}\otimes\mathcal{L}_{\eta_{\text{sg}}^{(2)},\Bar{n}_2})(\Psi_{AB}^{\pm})\Pi_{AB}\nonumber\\
		&\quad=\frac{1}{2}(x_1y_2+y_1x_2)\Phi_{AB}^+\nonumber\\
		&\qquad+\frac{1}{2}(x_1y_2+y_1x_2)\Phi_{AB}^-\nonumber\\
		&\qquad+\frac{1}{2}(x_1x_2+y_1y_2\pm z_1z_2)\Psi_{AB}^+\nonumber\\
		&\qquad+\frac{1}{2}(x_1x_2+y_1y_2\mp z_1z_2)\Psi_{AB}^-,
	\end{align}
	in the high-loss low-noise regime, and in the symmetric case $\eta_{\text{sg}}^{(1)}=\eta_{\text{sg}}^{(2)}=\eta_{\text{sg}}$ and $\Bar{n}_1=\Bar{n}_2=\Bar{n}$, we obtain
	\begin{align}
		&F(f_0)\nonumber\\
		&\quad\coloneqq\frac{\bra{\Phi^+}\Pi(\mathcal{L}_{\eta_{\text{sg}}^{(1)},\Bar{n}_1}\otimes\mathcal{L}_{\eta_{\text{sg}}^{(2)},\Bar{n}_2})(\rho^S(f_0))\Pi\ket{\Phi^+}}{\Tr[\Pi(\mathcal{L}_{\eta_{\text{sg}}^{(1)},\Bar{n}_1}\otimes\mathcal{L}_{\eta_{\text{sg}}^{(2)},\Bar{n}_2})(\rho^S(f_0))\Pi]}\\
		&\quad\approx \frac{1}{4}\left(1+\frac{4f_0-1}{\left(1+\frac{\Bar{n}}{\eta_{\text{sg}}}\right)^2}\right). \label{eq-fidelity_nonideal_bell_state}
	\end{align}
	Note that $1/4\leq F\leq f_0$. See Fig.~\ref{fig-fidelity_snr} for a plot of this fidelity as a function of the signal-to-noise ratio.
	
\subsubsection{Background photon flux}

	The background photon number $\Bar{n}$ can be expressed in terms of the photon flux/rate at a receiving site. Let $\mathcal{R}$ be the number of background photons per second detected at a receiving site and $\Delta T$ be the coincidence time-window. Then, $\Bar{n}=\mathcal{R}\Delta T$. Assuming background photons collected from, e.g., moonlight or sunlight, are the dominant source of noise, we have the following expression for the background photon rate \cite{Miao2005background,gruneisen2015sky_background}:
	\begin{equation}\label{eq-irradiance_flux}
		\mathcal{R}=\frac{H\Omega_\text{fov}\pi r^2\Delta\lambda}{hc/\lambda},
	\end{equation}
	where $hc/\lambda$ is the photon energy at mean wavelength $\lambda$ ($h$ is Planck's constant and $c$ is the speed of light), $\Delta\lambda$ is the filter bandwidth, $\Omega_\text{fov}$ is the field of view of a receiving telescope (in steradians, sr) with radius $r$, and $H$ is the total spectral irradiance in units $\text{W}\text{m}^{-2}\mu\text{m}^{-1}\text{sr}^{-1}$. In the case of daytime operating conditions, the total spectral irradiance includes direct solar irradiance as well as diffuse sky radiation, with the latter consisting mainly of solar light scattered by atmospheric constituents.
	
	The spectral irradiance is generally a complicated function of atmospheric conditions, the sun/moon sky position relative to the telescope pointing angle, time of day and year, etc. Thus, for simplicity, in what follows we keep $H$ as an open parameter but consider it to fall roughly within a typical range of $H\in[10^{-5},25]$ (in units $\text{W}\text{m}^{-2}\mu\text{m}^{-1}\text{sr}^{-1}$), associated with clear-sky conditions, with the lower value corresponding to a moonless clear night and the upper value corresponding to clear daytime conditions, when the sun is in near-view of the optical receiver (see, e.g., Refs.~\cite{Miao2005background,gruneisen2015sky_background}).
	
	\begin{figure}
    	\centering
    	\includegraphics[width=0.85\columnwidth]{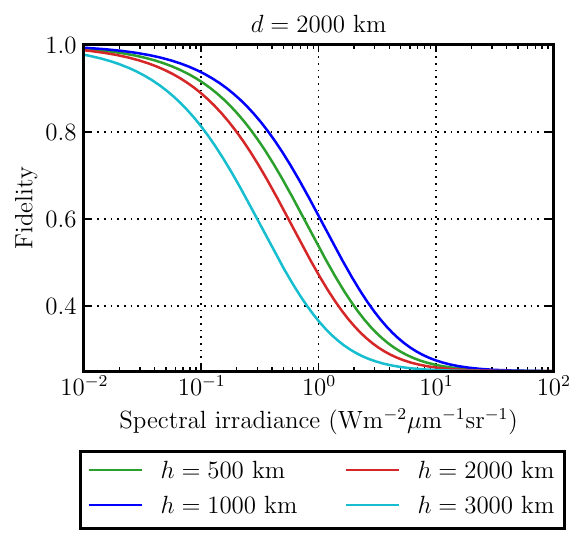}\\
    	\includegraphics[width=0.85\columnwidth]{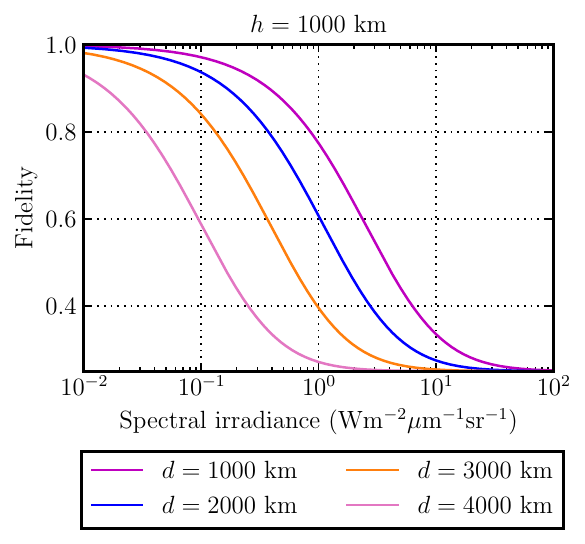}
    	\caption{Fidelity of satellite-to-ground entanglement transmission as a function of spectral irradiance. We consider transmission of the Bell state $\Phi^+$ according to the scenario depicted in Fig.~\ref{fig-atmosphere_geometry}. The fidelity is given in Eq.~\eqref{eq-fidelity_ideal}, and the average background photon number is given by $\Bar{n}=\mathcal{R}\Delta T$, with $\mathcal{R}$ given by Eq.~\eqref{eq-irradiance_flux}. In order to calculate $\mathcal{R}$, we let $\lambda=810~\text{nm}$, $\Delta\lambda=1~\text{nm}$, $\Omega_{\text{fov}}=100~\mu\text{sr}$, $r=0.5~\text{m}$, and $\Delta T=1~\text{ns}$; see, e.g., Refs.~\cite{gruneisen2015sky_background,LYL+2017,Ko+2018daylight}.} \label{fig-fidelity_v_H}
	\end{figure}

	Using the relation $\Bar{n}=\mathcal{R}\Delta T$, with $\mathcal{R}$ given by Eq.~\eqref{eq-irradiance_flux}, in Fig.~\ref{fig-fidelity_v_H} we plot the fidelity in Eq.~\eqref{eq-fidelity_ideal} as a function of the spectral irradiance $H$ for several orbital altitudes $h$ and ground-station separation distances $d$. To make the plot, we consider the situation depicted in Fig.~\ref{fig-atmosphere_geometry}, in which the satellite passes over the zenith of two ground stations and is at the midpoint between them. Note that spectral irradiance values on the order of  $1 \ \text{W}\text{m}^{-2}\mu\text{m}^{-1}\text{sr}^{-1}$ (and above) correspond to clear daytime conditions \cite{Miao2005background,Bon09, gruneisen2015sky_background}. Thus, for our chosen filter parameters, we see that entanglement distribution across, e.g., a ground-station separation distance of more than $2000$~km, only seems feasible during the night ($H\lesssim10^{-2} \ \text{W}\text{m}^{-2}\mu\text{m}^{-1}\text{sr}^{-1}$). We note, however, that these results are quite sensitive to the filtering parameters, owing to the steep slope of the fidelity in its mid-region.
	
	An interesting extension of these results would be to consider a dynamic model, in which one parameterizes the satellite-to-ground transmittance and background photon rate in time. We do such a parameterization for the transmittance in this work; however, parameterizing the background photon rate requires real-time modeling of, e.g., the sun position relative to the satellite orbit, modeling diffuse sky radiation, etc. Work along these lines has already been done for satellite-to-ground quantum key distribution between a satellite and a lone ground station (see, e.g., Ref.~\cite{gruneisen2015sky_background}). A full, dynamical analysis of the fidelity for a noisy, global-scale satellite-to-ground entanglement distribution protocol---utilizing, e.g., the asymmetric noise model derived above---is an interesting direction for future research.

\subsection{Simulation details}
	
	In order to perform our simulations and to obtain optimal satellite configurations, we used satellite constellations corresponding to the 42 pairs $(N_R,N_S)$ shown in Table~\ref{table-sat_configs}, where $N_R$ is the number of rings in the constellation and $N_S$ is the number of satellites per ring (see Fig.~\ref{fig-sat_architecture}).
	
	\begin{table}
		\renewcommand{\arraystretch}{1.3}
		\centering
		\begin{tabular}{|c c c|}
			\hline \multicolumn{3}{|c|}{$(N_R,N_S)$} \\ \hline\hline
			(2,10) & (4,8) & (5,8) \\
			(3,10) & (9,7) & (6,8) \\
			(4,5) & (8,7) & (7,8) \\
			(5,5) & (7,7) & (8,8) \\
			(6,5) & (6,7) & (9,8) \\
			(7,5) & (5,7) & (8,9) \\
			(8,5) & (4,7) & (9,9) \\
			(9,5) & (8,10) & (7,14) \\
			(4,6) & (9,10) & (7,15) \\
			(5,6) & (8,11) & (10,14) \\
			(6,6) & (10,10) & (10,15)  \\
			(7,6) & (4,13) & (15,15)  \\
			(8,6) & (5,13) & (16,16)  \\
			(9,6) & (7,13) & (20,20)  \\ \hline
		\end{tabular}
		\caption{Pairs $(N_R,N_S)$, consisting of the number $N_R$ of satellite rings and the number $N_S$ of satellites per ring, used in all of the simulations.}\label{table-sat_configs}
	\end{table}
	
	Let $\{\boldsymbol{r}_i(t)\in\mathbb{R}^3:1\leq i\leq N_RN_S\}$ be the positions of the satellites relative to the center of the earth at time $t$. If the satellites are at an altitude of $h$, then $\norm{\boldsymbol{r}_i(t)}_2=R_E+h$ for all $t$, where $R_E$ is the radius of the earth and $\norm{\cdot}_2$ denotes the Euclidean norm. Let $\boldsymbol{g}_j(t)\in\mathbb{R}^3$ be the position of the $j^{\text{th}}$ ground station relative to the center of the earth at time $t$.
	
	The distance between the $i^{\text{th}}$ satellite and the $j^{\text{th}}$ ground station at time $t$ is given by $L_{i,j}(t)=\norm{\boldsymbol{r}_i(t)-\boldsymbol{g}_j(t)}_2$. Then, the satellite-to-ground transmittance between the $i^{\text{th}}$ satellite, at altitude $h$, and the $j^{\text{th}}$ ground station is given at time $t$ by
	\begin{equation}
		\eta_{\text{sg}}^{(i,j)}(t;h)=\eta_{\text{fs}}(L_{i,j}(t))\eta_{\text{atm}}(L_{i,j}(t),h),
	\end{equation}
	with $\eta_{\text{fs}}(L_{i,j}(t))$ given by Eq.~\eqref{eq-fs_transmittance} and $\eta_{\text{atm}}(L_{i,j}(t),h)$ given by Eq.~\eqref{eq-atmospheric_transmittance_zenith}. The total transmittance $\eta_{\text{tot}}^{(i,j_1,j_2)}(t;h)$ at time $t$ corresponding to the $i^{\text{th}}$ satellite, at altitude $h$, transmitting one of a pair of entangled photons to ground station $j_1$ and the other photon to ground station $j_2$, is
	\begin{equation}\label{eq-eta_tot_pair}
		\eta_{\text{tot}}^{(i,j_1,j_2)}(t;h)=\eta_{\text{sg}}^{(i,j_1)}(t;h)\eta_{\text{sg}}^{(i,j_2)}(t;h).
	\end{equation}
	
	In order for a satellite to be considered within range of a given ground station pair, we require two conditions to be satisfied: 1) the satellite is visible to both ground stations; and 2) the total loss (given via Eq.~\eqref{eq-eta_tot_pair}) is less than 90 dB. If at least one of these two conditions is not satisfied, then the satellite is considered to be not within range of the ground station pair. Any interval of time in which at least one of the conditions is not satisfied is called a ``time gap''. We define the function
	\begin{multline}
		\text{range}(i,j_1,j_2,t)\\\coloneqq\left\{\begin{array}{l l} 1 & \text{if $j_1$ and $j_2$ visible to $i$ and}\\ & \text{$-10\log_{10}(\eta_{\text{tot}}^{(i,j_1,j_2)}(t;h))<90$ dB} \\[0.2cm] 0 & \text{otherwise,} \end{array}\right.
	\end{multline}
	which tells us whether the ground station pair $(j_1,j_2)$ is within range of the $i^{\text{th}}$ satellite, at altitude $h$, at time $t$.
	
	When performing our simulations, we find that at some times a satellite is within range of multiple ground station pairs. In other words, it can happen that a particular ground station pair is within range of multiple satellites at the same time. We anticipate that, in the near future, satellites will only have one entanglement source on board, so we impose the requirement that at any given time a satellite can distribute entanglement to only one ground-station pair. This requirement makes it necessary to uniquely assign a satellite to a ground-station pair at all times during the simulation. We assign a satellite to the ground-station pair that has the lowest loss among all ground-station pairs within range of that satellite. This type of assignment strategy means that, depending on the total number of satellites, there are times at which ground-station pairs do not receive any entangled photon pairs even though they are within range of a satellite (perhaps several), simply because the loss would be too high. More sophisticated time-sharing assignment strategies are possible, in which higher loss assignments are taken at certain times for the purpose of distributing entanglement to as many different ground-station pairs as possible. We do not consider such an assignment strategy here (see Ref.~\cite{PMD+20} for work in this direction), except for when there is a ground-station pair that has only one satellite in view, but that satellite is in range of several other ground stations. In this case, we assign that satellite to the ``lone'' ground-station pair even if the loss is higher than another possible assignment of that satellite. We let $s_t(j_1,j_2)\in\{1,2,\dotsc,N_RN_S\}$ denote the satellite assigned to the ground station pair $(j_1,j_2)$ at time $t$. For brevity, we write $s_t\equiv s_t(j_1,j_2)$ if the two ground stations being considered is clear from the context. If no satellite assignment exists for the pair $(j_1,j_2)$ at time $t$, then we set $s_t(j_1,j_2)=0$.
	
	For two ground stations $j_1$ and $j_2$ separated by a distance $d$, the average loss over a time $T$ for the satellite configuration given by $(N_R,N_S,h)$ is
	\begin{multline}
		\overline{\eta}_{\text{dB},T}^{(j_1,j_2)}(N_R,N_S,h,d)\\\coloneqq-10\log_{10}\left(\frac{1}{T}\sum_{t=1}^T \eta_{\text{tot}}^{(s_t,j_1,j_2)}(t;h)\right).
	\end{multline}
	If no satellite assignment exists for the pair $(j_1,j_2)$ at time $t$, then we set $\eta_{\text{tot}}^{(s_t,j_1,j_2)}(t;h)=0$. We write $\overline{\eta}_{\text{dB}}(N_R,N_S,h,d)$ when the ground station pair $(j_1,j_2)$ and the simulation time $T$ are understood from the context.
	
	We also define the average rate over time $T$ for two ground stations $j_1$ and $j_2$ separated by a distance $d$ for the satellite configuration given by $(N_R,N_S,h)$ as follows:
	\begin{equation}
		\overline{R}_T^{(j_1,j_2)}(N_R,N_S,h,d)\coloneqq \frac{1}{T}\sum_{t=1}^T \overline{P}^{(s_t,j_1,j_2)}(t),
	\end{equation}
	where
	\begin{equation}
		\overline{P}^{(s_t,j_1,j_2)}(t)\coloneqq R_{\text{source}}^{s_t}\eta_{\text{tot}}^{(s_t,j_1,j_2)}(t;h)
	\end{equation}
	is the average number of entangled pairs received by the ground stations $j_1$ and $j_2$ at time $t$, with $R_{\text{source}}^{s_t}$ being the source rate of the satellite $s_t$. For the simulations, we estimate $\overline{P}^{(s_t,j_1,j_2)}(t)$ in a single shot by taking a sample from the binomial distribution $\text{Bin}(n,p)$ with $n=R_{\text{source}}^{s_t}$ trials and success probability $p=\eta_{\text{tot}}^{(s_t,j_1,j_2)}(t;h)$ per trial. Throughout this work, we assume that $R_{\text{source}}^{s_t}=10^9$ ebits per second for all satellites \cite{CLZ+18}. We write $\overline{R}(N_R,N_S,h,d)$ when the ground station pair $(j_1,j_2)$ and the simulation time $T$ are understood from the context.

\begin{widetext}

\subsection{Figures of merit}

	Our figures of merit are based on entanglement distribution to a particular pair $(j_1,j_2)$ of ground stations. The goal is to determine the optimal satellite configuration such that there are no time gaps in a period of time $T$ for the pair $(j_1,j_2)$. The first three figures of merit separately optimize the total number of satellites, the average loss over the time $T$, and the average rate over the time $T$:
	\begin{align}
		N_{\text{opt},T}^{(j_1,j_2)}(h,d)&\coloneqq\left\{\begin{array}{l l} \text{minimum} & N_SN_R \\[0.2cm] \text{subject to} & \begin{minipage}[t]{10cm}\begin{itemize}[leftmargin=0.5cm] \item $\text{range}(s_t,j_1,j_2,t)=1~~\forall~1\leq t\leq T$ \item $\norm{\boldsymbol{r}_i(t)}_2=R_E+h~~\forall~1\leq i\leq N_RN_S,~1\leq t\leq T$ \item $j_1$, $j_2$ separated by distance $d$, \end{itemize}\end{minipage} \end{array}\right. \\[0.5cm]
		\overline{\eta}_{\text{dB},T}^{\text{opt},(j_1,j_2)}(h,d)&\coloneqq\left\{\begin{array}{l l} \text{minimum} & \overline{\eta}_{\text{dB},T}^{(j_1,j_2)}(N_R,N_S,h,d) \\[0.2cm] \text{subject to} & \begin{minipage}[t]{10cm}\begin{itemize}[leftmargin=0.5cm] \item $\text{range}(s_t,j_1,j_2,t)=1~~\forall~1\leq t\leq T$ \item $\norm{\boldsymbol{r}_i(t)}_2=R_E+h~~\forall~1\leq i\leq N_RN_S,~1\leq t\leq T$ \item $j_1$, $j_2$ separated by distance $d$, \end{itemize}\end{minipage} \end{array}\right. \\[0.5cm]
		\overline{R}_{T}^{\text{opt},(j_1,j_2)}(h,d)&\coloneqq\left\{\begin{array}{l l} \text{maximum} & \overline{R}_{T}^{(j_1,j_2)}(N_R,N_S,h,d) \\[0.2cm] \text{subject to} & \begin{minipage}[t]{10cm}\begin{itemize}[leftmargin=0.5cm] \item $\text{range}(s_t,j_1,j_2,t)=1~~\forall~1\leq t\leq T$ \item $\norm{\boldsymbol{r}_i(t)}_2=R_E+h~~\forall~1\leq i\leq N_RN_S,~1\leq t\leq T$ \item $j_1$, $j_2$ separated by distance $d$. \end{itemize}\end{minipage} \end{array}\right.
	\end{align}
	In all three cases, we optimize over the pairs $(N_R,N_S)$ shown in Table~\ref{table-sat_configs}. We write $N_{\text{opt}}(h,d)$, $\overline{\eta}_{\text{dB}}^{\text{opt}}(h,d)$, and $\overline{R}^{\text{opt}}(h,d)$ when both the ground station pair $(j_1,j_2)$ and the simulation time $T$ are understood from the context.
	
	In order to obtain a satellite configuration that balances both the total number of satellites and the average rate, we define the following figure of merit:
	\begin{equation}
		c_T^{(j_1,j_2)}(N_R,N_S,h,d)\coloneqq\frac{\overline{R}_T^{(j_1,j_2)}(N_R,N_S,h,d)}{N_RN_S},
	\end{equation}
	which has an intuitive interpretation as the average number of ebits per second per satellite in the network. From this, we define
	\begin{align}	
		C_{T}^{(j_1,j_2)}(h,d)&\coloneqq\left\{\begin{array}{l l} \text{maximum} & \overline{R}_T^{(j_1,j_2)}(N_R,N_S,h,d)/(N_RN_S) \\[0.2cm] \text{subject to} & \begin{minipage}[t]{10cm}\begin{itemize}[leftmargin=0.5cm] \item $\text{range}(s_t,j_1,j_2,t)=1~~\forall~1\leq t\leq T$ \item $\norm{\boldsymbol{r}_i(t)}_2=R_E+h~~\forall~1\leq i\leq N_RN_S,~1\leq t\leq T$ \item $j_1$, $j_2$ separated by distance $d$, \end{itemize}\end{minipage} \end{array}\right.
	\end{align}
	which is simply the figure of merit $c_T^{(j_1,j_2)}(N_R,N_S,h,d)$ optimized over the pairs $(N_R,N_S)$ in Table~\ref{table-sat_configs}. We write $C(h,d)$ when both the ground station pair $(j_1,j_2)$ and the simulation time $T$ are understood from the context.
	
	All of the quantities defined above can be defined in an analogous fashion for multiple ground station pairs (instead of just one ground station pair). For example, suppose that we have a set $\mathcal{V}$ of ground stations and a set $\mathcal{E}=\{(v,v',d_{v,v'}):v,v'\in\mathcal{V},d_{v,v'}\in\mathbb{R}\}$ telling us how the ground stations are connected pairwise to each other, with $d_{v,v'}$ being the physical distance between ground stations $v$ and $v'$. (Note that $(\mathcal{V},\mathcal{E})$ corresponds to a weighted graph, with $\mathcal{V}$ being the nodes of the graph and $\mathcal{E}$ being the weighed edges.) Then,
	\begin{equation}
		N_{\text{opt},T}^{\mathcal{V}}(h,\mathcal{E})\coloneqq\left\{\begin{array}{l l} \text{minimum} & N_RN_S \\[0.2cm] \text{subject to} & \begin{minipage}[t]{10cm}\begin{itemize}[leftmargin=0.5cm] \item $\text{range}(s_t,v,v',t)=1~~\forall~1\leq t\leq T,~~\forall~ v,v'\in\mathcal{V}$ \item $\norm{\boldsymbol{r}_i(t)}_2=R_E+h~~\forall~1\leq i\leq N_RN_S,~1\leq t\leq T$ \item $v$, $v'$ separated by distance $d_{v,v'}$, and $(v,v',d_{v,v'})\in\mathcal{E}$. \end{itemize}\end{minipage} \end{array}\right.
	\end{equation}
	Letting
	\begin{align}
		\overline{\eta}_{\text{dB},T}^{\mathcal{V}}(N_R,N_S,h,\mathcal{E})&\coloneqq-10\log_{10}\left(\frac{1}{|\mathcal{E}|T}\sum_{(v,v')\in\mathcal{E}}\sum_{t=1}^T\eta_{\text{tot}}^{(s_t,v,v')}(t;h)\right),\\
		\overline{R}_T^{\mathcal{V}}(N_R,N_S,h,\mathcal{E})&\coloneqq\frac{1}{|\mathcal{E}|T}\sum_{(v,v')\in\mathcal{E}}\sum_{t=1}^T R_{\text{source}}^{s_t}\eta_{\text{tot}}^{(s_t,v,v')}(t;h),
	\end{align}
	we can define the following quantities:
	\begin{align}
		\overline{\eta}_{\text{dB},T}^{\text{opt},\mathcal{V}}(h,\mathcal{E})&\coloneqq\left\{\begin{array}{l l} \text{minimum} & \overline{\eta}_{\text{dB},T}^{\mathcal{V}}(N_R,N_S,h,\mathcal{E}) \\[0.2cm] \text{subject to} & \begin{minipage}[t]{10cm}\begin{itemize}[leftmargin=0.5cm] \item $\text{range}(s_t,v,v',t)=1~~\forall~1\leq t\leq T,~~\forall~v,v'\in\mathcal{V}$ \item $\norm{\boldsymbol{r}_i(t)}_2=R_E+h~~\forall~1\leq i\leq N_RN_S,~1\leq t\leq T$ \item $v$, $v'$ separated by distance $d_{v,v'}$, and $(v,v',d_{v,v'})\in\mathcal{E}$, \end{itemize}\end{minipage} \end{array}\right. \\[0.5cm]
		\overline{R}_{T}^{\text{opt},\mathcal{V}}(h,\mathcal{E})&\coloneqq\left\{\begin{array}{l l} \text{maximum} & \overline{R}_{T}^{\mathcal{V}}(N_R,N_S,h,\mathcal{E}) \\[0.2cm] \text{subject to} & \begin{minipage}[t]{10cm}\begin{itemize}[leftmargin=0.5cm] \item $\text{range}(s_t,v,v',t)=1~~\forall~1\leq t\leq T,~~\forall~v,v'\in\mathcal{V}$ \item $\norm{\boldsymbol{r}_i(t)}_2=R_E+h~~\forall~1\leq i\leq N_RN_S,~1\leq t\leq T$ \item $v$, $v'$ separated by distance $d_{v,v'}$, and $(v,v',d_{v,v'})\in\mathcal{E}$, \end{itemize}\end{minipage} \end{array}\right. \\[0.5cm]
		C_{T}^{\mathcal{V}}(h,\mathcal{E})&\coloneqq\left\{\begin{array}{l l} \text{maximum} & \overline{R}_T^{\mathcal{V}}(N_R,N_S,h,\mathcal{E})/(N_RN_S) \\[0.2cm] \text{subject to} & \begin{minipage}[t]{10cm}\begin{itemize}[leftmargin=0.5cm] \item $\text{range}(s_t,v,v',t)=1~~\forall~1\leq t\leq T,~~\forall~v,v'\in\mathcal{V}$ \item $\norm{\boldsymbol{r}_i(t)}_2=R_E+h~~\forall~1\leq i\leq N_RN_S,~1\leq t\leq T$ \item $v$, $v'$ separated by distance $d_{v,v'}$, and $(v,v',d_{v,v'})\in\mathcal{E}$. \end{itemize}\end{minipage} \end{array}\right.
	\end{align}

\end{widetext}

\subsection{Quantum repeater rates}

\begin{figure}
		\centering
		\includegraphics[width=\columnwidth]{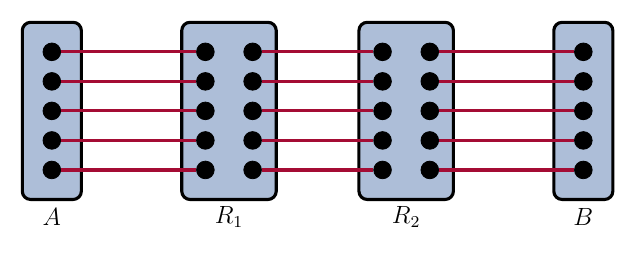}
		\caption{A repeater chain with $M=3$ elementary links. All of the elementary links have equal length, and there are $N_{\text{mem}}=5$ quantum memories per repeater half-node.}\label{fig-repeater_chain}
	\end{figure}

	In order to compare the satellite-based entanglement-distribution rates obtained in this work with rates that can be achieved using ground-based quantum repeater schemes, we consider a chain of quantum repeaters of total length $d$ in which there are $M$ elementary links and each repeater ``half-node'' has $N_{\text{mem}}$ quantum memories; see Fig.~\ref{fig-repeater_chain} for an example. This results in $N_{\text{mem}}$ parallel quantum repeater chains between the end nodes. If we allow entanglement distribution to occur independently for each of the parallel chains, and we assume that the quantum memories have infinite coherence time, then the expected number of time steps until one end-to-end pair is obtained, i.e., the expected waiting time, has been shown in \cite[Appendix~B]{KMSD19} to be
	\begin{equation}
		W_{N,N_{\text{mem}}}=\sum_{n=1}^{\infty}\left(1-\left(1-(1-p)^{n-1}\right)^M\right)^{N_{\text{mem}}}.
	\end{equation}
	Now, the duration of each time step, i.e., the repetition rate, is limited by the classical communication time between neighboring nodes for heralding of the signals. (This is the best-case scenario. We do not consider other factors that affect the repetition rate, such as the memory read-write time.) The classical communication time is given by $\frac{2(d/M)}{c}$, resulting in a repetition rate of $\frac{c}{2(d/M)}$ for each of the $N_{\text{mem}}$ parallel links of an elementary link. The total repetition rate is therefore $\frac{cN_{\text{mem}}}{2(d/M)}$. The formula in Eq.~\eqref{eq-rate_upper_bound} for the rate then follows.
	
	A higher rate than the one in Eq.~\eqref{eq-rate_upper_bound} can be achieved by allowing for spatial multiplexing, i.e., by allowing cross connections between the different parallel chains \cite{CJK07}. An analytic expression for the waiting time in this scenario, in the case of $M=2$ elementary links, has been derived in Ref.~\cite{BPL11}. A general formula for the waiting time for an arbitrary number $M$ of elementary links appears to be unknown.

\section{Acknowledgments}

	We dedicate this work to the memory of our mentor and guide Jonathan P. Dowling. Jon's unflinching support, his enthusiasm for research, and his inspiring vision for the future of quantum technologies and the quantum internet, made this work possible.

	SK acknowledges support from the National Science Foundation and the Natural Sciences and Engineering Research Council of Canada Postgraduate Scholarship. AJB acknowledges support from the National Science Foundation. RAD, MPB, and JPD acknowledge support from the Army Research Office, Air Force Office of Scientific Research and the National Science Foundation.

\section{Competing interests}

	The authors declare that there are no competing interests.

\section{Author contributions}

	SK, AJB, and JPD developed the main ideas behind the project. SK, RAD, and MPB developed and ran the simulations. SK and AJB wrote the manuscript. SK and AJB contributed equally to this work.

\section{Data and code availability}

	The datasets generated during the current study, and the code used to obtain the data, are available in the arXiv repository, \url{https://arxiv.org/abs/1912.06678}.

	
\bibliography{satellite}{}

\end{document}